\catcode`\@=11 
 
\def\nolabels{\def\wrlabel##1{}\def\eqlabel##1{}\def\reflabel##1{}}
\def\writelabels{\def\wrlabel##1{\leavevmode\vadjust{\rlap{\smash%
{\line{{\escapechar=` \hfill\rlap{\sevenrm\hskip.03in\string##1}}}}}}}%
\def\eqlabel##1{{\escapechar-1\rlap{\sevenrm\hskip.05in\string##1}}}%
\def\thlabel##1{{\escapechar-1\rlap{\sevenrm\hskip.05in\string##1}}}%
\def\reflabel##1{\noexpand\llap{\noexpand\sevenrm\string\string\string##1}}}
\nolabels
\global\newcount\secno \global\secno=0
\global\newcount\meqno \global\meqno=1
\global\newcount\mthno \global\mthno=1
\global\newcount\mexno \global\mexno=1
\global\newcount\mquno \global\mquno=1
\global\newcount\tblno \global\tblno=1
\def\newsec#1{\global\advance\secno by1 
\global\subsecno=0\xdef\secsym{\the\secno.}\global\meqno=1\global\mthno=1
\global\mexno=1\global\mquno=1\global\figno=1\global\tblno=1

\bigbreak\medskip\noindent{\bf\the\secno. #1}\writetoca{{\secsym} {#1}}
\par\nobreak\medskip\nobreak}
\xdef\secsym{}
\global\newcount\subsecno \global\subsecno=0
\def\subsec#1{\global\advance\subsecno by1 \global\subsubsecno=0
\xdef\subsecsym{\the\subsecno.}
\bigbreak\noindent{\bf\secsym\the\subsecno. #1}\writetoca{\string\quad
{\secsym\the\subsecno.} {#1}}\par\nobreak\medskip\nobreak}
\xdef\subsecsym{}
\global\newcount\subsubsecno \global\subsubsecno=0
\def\subsubsec#1{\global\advance\subsubsecno by1
\bigbreak\noindent{\it\secsym\the\subsecno.\the\subsubsecno.
                                   #1}\writetoca{\string\quad
{\the\secno.\the\subsecno.\the\subsubsecno.} {#1}}\par\nobreak\medskip\nobreak}
\global\newcount\appsubsecno \global\appsubsecno=0
\def\appsubsec#1{\global\advance\appsubsecno by1 \global\subsubsecno=0
\xdef\appsubsecsym{\the\appsubsecno.}
\bigbreak\noindent{\it\secsym\the\appsubsecno. #1}\writetoca{\string\quad
{\secsym\the\appsubsecno.} {#1}}\par\nobreak\medskip\nobreak}
\xdef\appsubsecsym{}
\def\appendix#1#2{\global\meqno=1\global\mthno=1\global\mexno=1
\global\figno=1\global\tblno=1
\global\subsecno=0\global\subsubsecno=0
\global\appsubsecno=0
\xdef\appname{#1}
\xdef\secsym{\hbox{#1.}}
\bigbreak\bigskip\noindent{\bf Appendix #1. #2}
\writetoca{Appendix {#1.} {#2}}\par\nobreak\medskip\nobreak}
%
%
\def\eqnn#1{\xdef #1{(\secsym\the\meqno)}\writedef{#1\leftbracket#1}%
\global\advance\meqno by1\wrlabel#1}
\def\eqna#1{\xdef #1##1{\hbox{$(\secsym\the\meqno##1)$}}
\writedef{#1\numbersign1\leftbracket#1{\numbersign1}}%
\global\advance\meqno by1\wrlabel{#1$\{\}$}}
\def\eqn#1#2{\xdef #1{(\secsym\the\meqno)}\writedef{#1\leftbracket#1}%
\global\advance\meqno by1$$#2\eqno#1\eqlabel#1$$}
%
%
\def\thm#1{\xdef #1{\secsym\the\mthno}\writedef{#1\leftbracket#1}%
\global\advance\mthno by1\wrlabel#1}
\def\exm#1{\xdef #1{\secsym\the\mexno}\writedef{#1\leftbracket#1}%
\global\advance\mexno by1\wrlabel#1}
%
%
\def\tbl#1{\xdef #1{\secsym\the\tblno}\writedef{#1\leftbracket#1}%
\global\advance\tblno by1\wrlabel#1}
%
\newskip\footskip\footskip14pt plus 1pt minus 1pt 
\def\f@@t{\baselineskip\footskip\bgroup\aftergroup\@foot\let\next}
\setbox\strutbox=\hbox{\vrule height9.5pt depth4.5pt width0pt}
\global\newcount\ftno \global\ftno=0
\def\foot{\global\advance\ftno by1\footnote{$^{\the\ftno}$}}
%
\newwrite\ftfile
\def\footend{\def\foot{\global\advance\ftno by1\chardef\wfile=\ftfile
$^{\the\ftno}$\ifnum\ftno=1\immediate\openout\ftfile=foots.tmp\fi%
\immediate\write\ftfile{\noexpand\smallskip%
\noexpand\item{f\the\ftno:\ }\pctsign}\findarg}%
\def\footatend{\vfill\eject\immediate\closeout\ftfile{\parindent=20pt
\centerline{\bf Footnotes}\nobreak\bigskip\input foots.tmp }}}
\def\footatend{}
%
%
\global\newcount\refno \global\refno=1
\newwrite\rfile
\def\ref{\the\refno\nref}
\def\bref{\nref}
\def\nref#1{\xdef#1{\the\refno}\writedef{#1\leftbracket#1}%
\ifnum\refno=1\immediate\openout\rfile=refs.tmp\fi
\global\advance\refno by1\chardef\wfile=\rfile\immediate
\write\rfile{\noexpand\item{[#1]\ }\reflabel{#1\hskip.31in}\pctsign}\findarg}
\def\findarg#1#{\begingroup\obeylines\newlinechar=`\^^M\pass@rg}
{\obeylines\gdef\pass@rg#1{\writ@line\relax #1^^M\hbox{}^^M}%
\gdef\writ@line#1^^M{\expandafter\toks0\expandafter{\striprel@x #1}%
\edef\next{\the\toks0}\ifx\next\em@rk\let\next=\endgroup\else\ifx\next\empty%
\else\immediate\write\wfile{\the\toks0}\fi\let\next=\writ@line\fi\next\relax}}
\def\striprel@x#1{} \def\em@rk{\hbox{}}
\def\lref{\begingroup\obeylines\lr@f}
\def\lr@f#1#2{\gdef#1{\ref#1{#2}}\endgroup\unskip}

\def\addref#1{\immediate\write\rfile{\noexpand\item{}#1}} 
\def\startrefs#1{\immediate\openout\rfile=refs.tmp\refno=#1}
\def\xref{\expandafter\xr@f}\def\xr@f[#1]{#1}
\def\refs#1{[\r@fs #1{\hbox{}}]}
\def\r@fs#1{\edef\next{#1}\ifx\next\em@rk\def\next{}\else
\ifx\next#1\xref #1\else#1\fi\let\next=\r@fs\fi\next}
%

%
 \newwrite\ffile\global\newcount\figno \global\figno=1
%
%
\def\fig{\the\figno\nfig}
\def\nfig#1{\xdef#1{\secsym\the\figno}%
\writedef{#1\leftbracket \noexpand~\the\figno}%
\ifnum\figno=1\immediate\openout\ffile=figs.tmp\fi\chardef\wfile=\ffile%
\immediate\write\ffile{\noexpand\medskip\noexpand\item{Figure\ \the\figno. }
\reflabel{#1\hskip.55in}\pctsign}\global\advance\figno by1\findarg}
\def\xfig{\expandafter\xf@g}\def\xf@g \penalty\@M\ {}
\def\figs#1{figs.~\f@gs #1{\hbox{}}}
\def\f@gs#1{\edef\next{#1}\ifx\next\em@rk\def\next{}\else
\ifx\next#1\xfig #1\else#1\fi\let\next=\f@gs\fi\next}
%
%
\newwrite\lfile

{\escapechar-1\xdef\pctsign{\string\%}\xdef\leftbracket{\string\{}
\xdef\rightbracket{\string\}}\xdef\numbersign{\string\#}}

\def\writestop{\def\writestoppt{\immediate\write\lfile{\string\pageno%
\the\pageno\string\startrefs\leftbracket\the\refno\rightbracket%
\string\def\string\secsym\leftbracket\secsym\rightbracket%
\string\secno\the\secno\string\meqno\the\meqno}\immediate\closeout\lfile}}
\def\writestoppt{}\def\writedef#1{}

\def\seclab#1{\xdef #1{\the\secno}\writedef{#1\leftbracket#1}\wrlabel{#1=#1}}

\def\subseclab#1{\xdef #1{\secsym\the\subsecno}%
\writedef{#1\leftbracket#1}\wrlabel{#1=#1}}
\def\appsubseclab#1{\xdef #1{\secsym\the\appsubsecno}%
\writedef{#1\leftbracket#1}\wrlabel{#1=#1}}
\def\subsubseclab#1{\xdef #1{\secsym\the\subsecno.\the\subsubsecno}%
\writedef{#1\leftbracket#1}\wrlabel{#1=#1}}
\newwrite\tfile \def\writetoca#1{}
\def\leaderfill{\leaders\hbox to 1em{\hss.\hss}\hfill}
\def\writetoc{\immediate\openout\tfile=toc.tmp
   \def\writetoca##1{{\edef\next{\write\tfile{\noindent ##1
   \string\leaderfill {\noexpand\number\pageno} \par}}\next}}}

%
\catcode`\@=12 
%
%
%
%
%
\def\dbend{{\manual\char127}}
\def\d@nger{\medbreak\begingroup\clubpenalty=10000
    \def\par{\endgraf\endgroup\medbreak} \noindent\hang\hangafter=-2
    \hbox to0pt{\hskip-\hangindent\dbend\hfill}\ninepoint}
\outer\def\danger{\d@nger}

\def\p{\partial}

\def\darr#1{\raise1.5ex\hbox{$\leftrightarrow$}\mkern-16.5mu #1}
\def\half{{\textstyle{1\over2}}} 

%
%
\def\al{\alpha}
\def\be{\beta}
  
\def\de{\delta}  \def\De{\Delta}
\def\ep{\epsilon}  
\def\ze{\zeta}
\def\et{\eta}

\def\la{\lambda} \def\La{\Lambda}
\def\rh{\rho}
\def\si{\sigma}  
\def\ta{\tau}
  
\def\ph{\phi}    \def\vph{\varphi}

%
%

%

%
%
\def\cA{{\cal A}}

\def\cI{{\cal I}}

\def\cO{{\cal O}}

 \def\cS{{\cal S}}

\def\cW{{\cal W}}  

%

%

%
%
\def\amsyes{y }

\def\answ{y }

\ifx\answ\amsyes
\input amssym.def


\def\CC{{\Bbb C}}
\def\ZZ{{\Bbb Z}}

\def\bfg{{\frak g}}

\def\hg{{\widehat{\frak g}}}
\def\whg{\hg}
\else
\def\ZZ{{Z\!\!\!Z}}              
\def\CC{{I\!\!\!\!C}}

\def\bfg{{\bf g}}

\def\hg{\hat{\bf g}}

\def\cA{{\cal A}} 

\fi
%

%
%

%
%

\newsymbol\ltimes 226E
\newsymbol\rtimes 226F
%
%
%

\def\CMP#1{Comm.\ Math.\ Phys.\ {\bf #1}}

\def\FAP#1{Funct.\ Anal.\ Appl.\ {\bf #1}}
\def\IJMP#1{Int.\ J.\ Mod.\ Phys.\ {\bf #1}}

\def\JMP#1{J.\ Math.\ Phys.\ {\bf #1}}
\def\JPA#1{J.\ Phys.\ {\bf A{#1}}}

\def\JSP#1{J.\ Stat.\ Phys. {\bf {#1}}}

\def\LMP#1{Lett.\ Math.\ Phys.\ {\bf #1}}

\def\MPL#1{Mod.\ Phys.\ Lett.\ {\bf #1}}
\def\NPB#1{Nucl.\ Phys.\ {\bf B#1}}
\def\PLB#1{Phys.\ Lett.\ {\bf {#1}B}}

\def\PRep#1{Phys.\ Rep.\ {\bf #1}}

\def\PRB#1{Phys.\ Rev.\ {\bf B{#1}}}

\def\PRL#1{Phys.\ Rev.\ Lett.\ {\bf #1}}

%

%
%
\def\SMu{\hbox{\lower 3pt\hbox{ \epsffile{su10.eps}}}}
\def\SMs{\hbox{\lower 3pt\hbox{ \epsffile{ss10.eps}}}}
\def\SMd{\hbox{\lower 3pt\hbox{ \epsffile{sd10.eps}}}}

\def\SMS{\leavevmode\vadjust{\rlap{\smash%
{\line{{\escapechar=` \hfill\rlap{\hskip.3in%
                 \hbox{\lower 2pt\hbox{\epsffile{sd10.eps}}}}}}}}}}
\def\SMH{\leavevmode\vadjust{\rlap{\smash%
{\line{{\escapechar=` \hfill\rlap{\hskip.3in%
                 \hbox{\lower 2pt\hbox{\epsffile{su10.eps}}}}}}}}}}
%
%
\def\LW#1{\lower .5pt \hbox{$\scriptstyle #1$}}
\def\LWr#1{\lower 1.5pt \hbox{$\scriptstyle #1$}}
\def\LWrr#1{\lower 2pt \hbox{$\scriptstyle #1$}}
\def\RSr#1{\raise 1pt \hbox{$\scriptstyle #1$}}

\def\txt#1{$\hbox{$#1$}$}
  
 
\magnification=1200

\vsize=22.5truecm  \hsize=16.5truecm

\nopagenumbers
\hfuzz=40pt
\pageno=0
\def\onth{{\scriptstyle{1\over3}}} \def\thqu{{\scriptstyle{3\over4}}}
\def\twth{{\scriptstyle{2\over3}}} \def\thha{{\scriptstyle{3\over2}}}
\def\fiqu{{\scriptstyle{5\over4}}} \def\onqu{{\scriptstyle{1\over4}}}
\def\half{{\scriptstyle{1\over2}}}
\def\scr#1{{\scriptstyle{#1}}} \def\txt#1{{\textstyle{#1}}}
\def\eqv{~\equiv~} \def\eql{~=~} \def\ceff{c_{\rm eff}}
\def\qn#1{(q)_{#1}} \def\qeii{QEI$\!$I}
\def\id{{\bf 1}} \def\bm{{\bf m}} \def\bu{{\bf u}} \def\bM{{\bf M}}
\def\bin#1#2{\left( \matrix{ #1 \cr #2 \cr} \right)}
\def\qbin#1#2{\left[ \matrix{ #1 \cr #2 \cr} \right]}
\def\cS{{\cal S}} \def\cSph{\cS^{\rm ph}}
\def\cSps{\cS^{\rm ps}} \def\bG{{\bf G}} \def\bA{{\bf A}} \def\bB{{\bf B}}
\def\PLA#1{Phys.\ Lett.\ {\bf A{#1}}} 
\def\PTPS#1{Prog.\ Theor.\ Phys.\ Suppl.\ {\bf {#1}}}
\def\wh#1{\widehat{#1}} 
\def\hb{\hfill\break}
\def\wt#1{\widetilde{#1}} \def\latot{\la_{\rm tot}}
%

\bref\Ha{
D.~Haldane, 
\PRL{67} (1991) 937.}

\bref\Wuetal{
Y.-S.~Wu, 
\PRL{73} (1994) 922; \hb
C.~Nayak and F.~Wilczek, \PRL{73} (1994) 2740; \hb
S.B.~Isakov, \MPL{B8} (1994) 319; \hb
A.~Dasni\`eres de Veigy and S.~Ouvry, \PRL{72} (1994) 600; {\it ibid.},
\MPL{B9} (1995) 271, [{\tt cond-mat/9411036}]; \hb
A.K.~Rajagopal, \PRL{74} (1995) 1048; \hb
Y.-S.~Wu and Y.~Yu, \PRL{75} (1995) 890; \hb
S.B.~Isakov, D.P.~Arovas, J.~Myrheim, A.P.~Polychronakos, 
\PLA{212} (1996) 299, [{\tt cond-mat/9601108}].}

\bref\FK{
T.~Fukui and N.~Kawakami, 
\PRB{51} (1995) 5239, [{\tt cond-mat/9408015}].}

\bref\Hab{
F.~Haldane, \PRL{66} (1991) 1529; \hb
F.~Haldane, Z.~Ha, J.~Talstra, D.~Bernard and V.~Pasquier,
\PRL{69} (1992) 2021.}

\bref\BPS{
D.~Bernard, V.~Pasquier and D.~Serban,
\NPB{428} (1994) 612, [{\tt hep-th/9404050}].}

\bref\BLSa{
P.~Bouwknegt, A.~Ludwig and K.~Schoutens, 
\PLB{338} (1994) 448, [{\tt hep-th/9406020}].}

\bref\BLSb{
P.~Bouwknegt, A.~Ludwig and K.~Schoutens, 
\PLB{359} (1995) 304, [{\tt hep-th/9412108}].}

\bref\Tak{
L.~Takhtajan, \PLA{87} (1982) 479; \hb
L.~Faddeev and N.~Reshetikhin, Ann.\ Phys.\ {\bf 167} (1986) 227; \hb
I.~Affleck and F.~Haldane, 
\PRB{36} (1987) 5291; \hb
N.~Reshetikhin, \JPA{24} (1991) 3299; \hb
P.~Fendley, \PRL{71} (1993) 2845, [{\tt cond-mat/9304031}].}

\bref\Sca{
K.~Schoutens, 
\PRL{79} (1997) 2608, [{\tt cond-mat/9706166}].}

\bref\Scb{
K.~Schoutens, 
\PRL{81} (1998) 15704, [{\tt cond-mat/9803169}].}

\bref\FrS{
H.~Frahm and M.~Stahlsmeier, 
\PLA{250} (1998) 293, [{\tt cond-mat/9803381}].}

\bref\Gai{
J.~Gaite, \NPB{525} (1998) 627, [{\tt hep-th/9804025}].}

\bref\ES{
R.~van Elburg and K.~Schoutens, \PRB{58} (1998) 15704, 
[{\tt cond-mat/9801272}].}

\bref\BSc{
P.~Bouwknegt and K.~Schoutens, {\it Exclusion statistics in conformal field
theory -- generalized fermions and spinons for level-$1$ WZW theories},
\NPB{}, to appear, [{\tt hep-th/9810113}].}

\bref\GS{
S.~Guruswamy and K.~Schoutens, {\it Non-abelian exclusion statistics},
[{\tt cond-mat/9903045}].}

\bref\KM{
R.~Kedem and  B.M.~McCoy, \JSP{71} (1994) 865, [{\tt hep-th/9210146}];\hb
G.~Albertini, S.~Dasmahapatra and B.~McCoy,
\IJMP{A7} Suppl.\ {\bf 1A} (1992) 1, ; {\it ibid.},
\PLA{170} (1992) 397; \hb
S.~Dasmahapatra, R.~Kedem, B.M.~McCoy and E.~Melzer,
\JSP{74} (1994) 239, [{\tt hep-th/9304150}].}

\bref\BPW{
D.~O'Brien, P.~Pearce and S.O.~Warnaar, \NPB{501} (1997) 773.}

\bref\BLZ{
V.~Bazhanov, S.~Lukyanov and A.~Zamolodchikov, 
\CMP{177} (1996) 381, [{\tt hep-th/9412229}]; {\it ibid.},
\CMP{190} (1997) 247,  [{\tt hep-th/9604044}].}

\bref\ABZ{
A.B.~Zamolodchikov, Advanced Studies in Pure Math.\ {\bf 19} (1989) 641; \hb
D.~Bernard and A.~Le Clair, \NPB{340} (1990) 721; \hb 
C.~Ahn, D.~Bernard and A.~Le Clair, \NPB{346} (1990) 409; \hb
F.A.~Smirnov, \NPB{337} (1990) 156; {\it ibid.},
\IJMP{A6} (1991) 1407; \hb
G.~Mussardo, \PRep{218} (1992) 215.}

\bref\BMP{
A.~Berkovich, B.~McCoy and P.~Pearce, \NPB{519} (1998) 597; \hb
A.~Berkovich and B.~McCoy, {\it The perturbation $\vph_{2,1}$ of the 
$M(p,p+1)$ models of conformal field theory and related polynomial
character identities}, [{\tt math.QA/9809066}]; \hb
S.O.~Warnaar, {\it $q$-Trinomial identities}, [{\tt math.QA/9810018}].}

\bref\LP{
J.~Lepowski and M.~Primc, {\it Structure of the standard modules for
the affine Lie algebra $A_1^{(1)}$}, Contemp.\ Math.\ {\bf 46},
(Amer.\ Math.\ Soc., Providence, 1985).}

\bref\KKMM{
R.~Kedem, T.R.~Klassen, B.M.~McCoy and E.~Melzer,
\PLB{304} (1993) 263, [{\tt hep-th/9211102}]; 
{\it ibid.},
\PLB{307} (1993) 68, [{\tt hep-th/9301046}];\hb
S.~Dasmahapatra, R.~Kedem, T.R.~Klassen, B.M.~McCoy 
and E.~Melzer, \IJMP{B7} (1993) 3617, [{\tt hep-th/9303013}].}

\bref\Mel{
E.~Melzer, \IJMP{A9} (1994) 1115, [{\tt hep-th/9305114}]; \hb
A.~Berkovich, \NPB{431} (1994) 315, [{\tt hep-th/9403073}]; \hb
A.~Berkovich and B.M.~McCoy, \LMP{37} (1996) 49, [{\tt hep-th/9412030}]; \hb
S.O.~Warnaar, \JSP{82} (1996) 657, [{\tt hep-th/9501134}]; 
{\it ibid.}, \JSP{84} (1996) 49, [{\tt hep-th/9508079}]; \hb
A.~Berkovich, B.M.~McCoy and A.~Schilling, 
\CMP{191} (1998) 325, [{\tt q-alg/9607020}].}

\bref\FQW{
O.~Foda and Y.-H.~Quano, 
\IJMP{A12} (1997) 1651, [{\tt hep-th/9408086}]; \hb
O.~Foda, K.S.M.~Lee and T.A.~Welsh,
\IJMP{A13} (1998) 4967, [{\tt q-alg/9710025}]; \hb  
O.~Foda and T.~Welsh, {\it Melzer's identities revisited}, 
[{\tt math.QA/9811156}].}

\bref\ScW{
A.~Schilling, \NPB{459} (1996) 393, [{\tt hep-th/9508050}]; \hb
A.~Schilling and S.O.~Warnaar,
The Ramanujan Journal {\bf 2} (1998) 459, [{\tt q-alg/9701007}].}

\bref\FSt{
B.L.~Feigin and A.V.~Stoyanovsky, {\it Quasi-particle models for the 
representations of Lie algebras and geometry of flag manifolds}, 
[{\tt hep-th/9308079}]; {\it ibid.}, \FAP{28} (1994) 55.}

\bref\Geo{
G.~Georgiev, 
J.\ Pure Appl.\ Algebra {\bf 112} (1996) 247, [{\tt hep-th/9412054}]; 
{\it ibid.}, {\it Combinatorial constructions 
of modules for infinite-dimensional Lie algebras, II.\ Parafermionic space}, 
[{\tt q-alg/9504024}].}

\bref\BSa{
P.~Bouwknegt and K.~Schoutens, 
\NPB{482} (1996) 345, [{\tt hep-th/9607064}].}

\bref\BM{
A.~Berkovich and B.~McCoy, {\it The universal chiral partition function 
for exclusion statistics}, [{\tt hep-th/9808013}].}

\bref\KMM{
R.~Kedem, B.~McCoy and E.~Melzer, in ``Recent progress in Statistical
Mechanics and Quantum Field Theory'', eds.\ P.~Bouwknegt et al.,
(World Scientific, Singapore, 1995), [{\tt hep-th/9304056}].}

\bref\BW{
D.~Bernard and Y.-S.~Wu, {\it A note on statistical interactions and
the thermodynamic Bethe Ansatz}, in ``New developments in integrable
systems and long-range interaction models'', Nankai Lecture Notes on
Mathematical Physics, (World Scientific, Singapore, 1994),
[{\tt cond-mat/9404025}].}

\bref\Hik{
K.~Hikami, 
\PLA{205} (1995) 364.}

\bref\BH{
P.~Bouwknegt, {\it $q$-Identities and affinized 
projective varieties, I.\ Quadratic monomial ideals}, 
[{\tt math-ph/9902010}]; \hb
P.~Bouwknegt and N.~Halmagyi, {\it $q$-Identities and affinized 
projective varieties, II.\ Flag varieties}, [{\tt math-ph/9903033}].}

\bref\BFa{
A.~Bytsko and A.~Fring, 
\NPB{532} (1998) 588, [{\tt hep-th/9803005}].}

\bref\YY{
C.N.~Yang and C.O.~Yang, \JMP{10} (1969) 1115; \hb
B.~Sutherland, \PRL{20} (1967) 98.} 
 
\bref\AlBZa{
Al.B.~Zamolodchikov, \NPB{358} (1991) 497; {\it ibid.},
\NPB{358} (1991) 524; {\it ibid.}, \NPB{366} (1991) 122; \hb
P.~Fendley, H.~Saleur and Al.B.~Zamolodchikov,
\IJMP{A8} (1993) 5751.} 

\bref\RTV{
F.~Ravanini, \PLB{282} (1992) 73, [{\tt hep-th/9202020}]; \hb
F.R.~Ravanini, R.~Tateo and A.~Valleriani,
\IJMP{A8} (1993) 1707, [{\tt hep-th/9207040}].}

\bref\BR{
V.V.~Bazhanov and N.Y.~Reshetikhin, \IJMP{A4} (1989) 115;
{\it ibid.}, \JPA{23} (1990) 1477; {\it ibid.}, \PTPS{102} (1990) 301.}

\bref\AlBZb{
Al.B.~Zamolodchikov, \NPB{342} (1990) 695.}

\bref\KlM{
T.~Klassen and E.~Melzer,
\NPB{338} (1990) 485; {\it ibid.},
\NPB{350} (1991) 635; {\it ibid.},
\NPB{370} (1992) 511.}
 
\bref\NYa{
A.~Nakayashiki and Y.~Yamada, \CMP{178} (1996) 179, 
[{\tt hep-th/9504052}].}

\bref\BFb{
A.~Bytsko and A.~Fring, {\it Factorized combinations of Virasoro 
characters}, [{\tt hep-th/9809001}].}

\bref\NRT{
W.~Nahm, A.~Recknagel and M.~Terhoeven, \MPL{A8} (1993) 1835,
[{\tt hep-th/9211034}].}

\bref\DKKMM{
S.~Dasmahapatra, R.~Kedem, T.~Klassen, B.~McCoy and E.~Melzer,
\IJMP{B7} (1993) 3617, [{\tt hep-th/9303013}].}

\bref\RS{
B.~Richmond and G.~Szekeres, J.\ Austral.\ Math.\ Soc.\ (Series A)
{\bf 31} (1981) 362.}

\bref\Kir{
A.~Kirillov, 
\PTPS{118} (1995) 61, [{\tt hep-th/9408113}].}

\bref\KN{
A.~Kuniba and T.~Nakanishi, 
\MPL{A7} (1992) 3487, [{\tt hep-th/9206034}]; \hb
A.~Kuniba, T.~Nakanishi and J.~Suzuki, 
\MPL{A8} (1993) 1649, [{\tt hep-th/9301018}].}

\bref\Sub{
J.~Suzuki, 
\JPA{31} (1998) 6887, [{\tt cond-mat/9805242}].}

\bref\Sua{
J.~Suzuki, {\it Spinons in magnetic chains of arbitrary spins at finite
temperatures}, [{\tt cond-mat/9807076}].}

\bref\BSd{
P.~Bouwknegt and K.~Schoutens, {\it Spinon decomposition and 
Yangian structure of $\widehat{\frak{sl}_n}$ modules}, in
``Geometric Analysis and Lie Theory in Mathematics and
Physics, Australian Mathematical Society Lecture Series {\bf 11}, 
{\it eds.} A.L.~Carey and M.K.~Murray, (Cambridge University Press,
Cambridge, 1997), [{\tt q-alg/9703021}].}

\bref\BMS{
A.~Berkovich, B.~McCoy and A.~Schilling, \CMP{191} (1998) 325, 
[{\tt q-alg/9607020}].}

\bref\HKKOTY{
G.~Hatayama, A.~Kirillov, A.~Kuniba, M.~Okado, T.~Takagi and Y.~Yamada,
\NPB{536} (1999) 575, [{\tt math.QA/9802085}]; \hb 
G.~Hatayama, A.~Kuniba, M.~Okado, T.~Takagi and Y.~Yamada,
{\it Remarks on fermionic formula}, [{\tt math.QA/9812022}].}

\bref\GNOS{
P.~Goddard, W.~Nahm, D.~Olive and A.~Schwimmer, \CMP{107} (1986) 179;\hb
D.~Bernard and J.~Thierry-Mieg, \CMP{111} (1987) 181.}

\bref\NYc{
A.~Nakayashiki and Y.~Yamada, {\it On spinon character formulas}, in
``Frontiers in Quantum Field Theories'', {\it eds.\ }
H.~Itoyama et al., (World Scientific, Singapore, 1996).}

\bref\Ya{
Y.~Yamada, {\it On $q$-Clebsch Gordan rules and the spinon character 
formulas for affine $C_2^{(1)}$ algebra}, [{\tt q-alg/9702019}].}

\bref\GR{
G.~Gasper and M.~Rahman, {\it Basic hypergeometric series}, 
Encycl.\ of Math.\ and its Appl.\ {\bf 35} (Cambridge University Press,
Cambridge, 1990).}

\bref\KR{
A.~Kirillov and N.~Reshetikhin,
J.\ Sov.\ Math.\ {\bf 41} (1988) 925.}

\bref\NYb{
A.~Nakayashiki and Y.~Yamada,
Selecta Math.\ (N.S.) {\bf 3} (1997) 547, [{\tt q-alg/9512027}].}

\bref\FKY{
T.~Fukui, N.~Kawakami and S.-K.~Yang, 
{\it Spin-$S$ generalization of fractional exclusion statistics}, 
[{\tt cond-mat/9507143}].}

\bref\ANOT{
T.~Arakawa, T.~Nakanishi, K.~Oshima and A.~Tsuchiya, \CMP{181} (1996)
157, [{\tt q-alg/9507025}].}

\bref\BSb{
P.~Bouwknegt and K.~Schoutens, {\it Non-abelian electrons: $SO(5)$ superspin
regimes for correlated electrons on a two-leg ladder}, 
\PRL{}, to appear, [{\tt cond-mat/9805232}].}


%
%
\line{}
\vskip2cm
\centerline{\bf EXCLUSION STATISTICS IN CONFORMAL FIELD THEORY}\smallskip
\centerline{\bf AND THE UCPF FOR WZW MODELS}\bigskip
\vskip1cm

\centerline{Peter BOUWKNEGT, Leung CHIM and David RIDOUT%
\footnote{$^\dagger$}{Present address: Department of Mathematics, 
University of Western Australia, Nedlands WA~6907.}}
\bigskip

\centerline{\sl Department of Physics and Mathematical Physics}
\centerline{\sl University of Adelaide}
\centerline{\sl Adelaide, SA~5005, AUSTRALIA}
\bigskip
\vskip1cm

\centerline{\it Dedicated to the memory of Prof.\ H.S.\ Green}\vskip1.5cm

\centerline{\bf ABSTRACT}\bigskip

\vbox{\leftskip 2.0truecm \rightskip 2.0truecm
\noindent In this paper we further elaborate on the notion of 
fractional exclusion statistics, as introduced by Haldane,  
in two-dimensional conformal field theory, and its
connection to the Universal Chiral Partition Function as defined by
McCoy and collaborators.  We will argue that in general, besides the
pseudo-particles introduced recently by Guruswamy and Schoutens, one
needs additional `null quasi-particles' to account for the null-states
in the quasi-particle Fock space.  We illustrate this in several
examples of WZW-models. 
}

\vfil
\leftline{ADP-99-12/M78}
\line{{\tt hep-th/9903176}\hfil March 1999}

\eject


\baselineskip=1.4\baselineskip

\footline{\hss \tenrm -- \folio\ -- \hss}

\newsec{Introduction}

Elementary excitations in low dimensional quantum many-body systems
may exhibit `fractional statistics' generalizing the usual Bose and
Fermi statistics. In such cases the single particle states available
to an excitation may depend on the entire particle content of the
multi-particle
state.  To handle such systems Haldane [\Ha] introduced a particular
form of, so-called, `fractional exclusion statistics' where the
statistical interactions are encoded into a matrix $G_{ab}$.  The
thermodynamics of an ideal gas of particles satisfying Haldane's
fractional exclusion statistics was subsequently worked out in a
series of papers [\Wuetal,\FK].

Fractional exclusion statistics arises naturally in quasi-particle
descriptions of two-dimensional Conformal Field Theories (CFTs).  Here
quasi-particles correspond to intertwiners (Chiral Vertex Operators or
CVOs) between the various representations of the chiral algebra and
the (chiral) spectrum is constructed by repeated application of the
modes of a preferred set of CVOs on the vacuum.  Inspired by [\Hab,\BPS],
such a basis was
first constructed for the $(\wh{\frak{sl}_2})_{k\geq1}$ WZW models
[\BLSa,\BLSb] in terms of a $j=1/2$ spinon field. This basis
in particular illuminates the appearance of fundamental spinons in 
spin-$S$ integrable spin chains whose effective CFT is an
$(\wh{\frak{sl}_2})_{k=2S}$ WZW model [\Tak].  

The virtue of the quasi-particle approach 
to CFT is that there is a simple method,
developed in [\Sca], to compute the thermodynamical properties of the
quasi-particles and expose their `fractional exclusion
statistics'. This method involves truncating the quasi-particle basis
in momentum space and finding recursion relations for the associated
finitized chiral characters.  {}From the recursion relations
one immediately derives the (total) single-level grand partition
function $\latot$ for the quasi-particles and hence their statistical
properties.  A large number of examples were subsequently worked out
[\Sca--\BSc].  
Particularly interesting applications of this approach
include the study of the fractional exclusion statistics of the edge
quasi-particle excitations over abelian quantum Hall states [\ES].  

In many cases, including $(\wh{\frak{sl}_2})_{k=1}$ WZW models 
[\Ha,\Sca] and
$\ZZ_k$ parafermions [\Gai,\BSc], it was discovered that the exclusion
statistics of these CFT quasi-particles is indeed of the type
introduced by Haldane.  All these examples involved quasi-particles
with abelian braid statistics, corresponding in the CFT to
intertwiners (CVOs) with a unique fusion path.  The corresponding
statistics is referred to as abelian exclusion statistics.  
At the same time it was obvious that quasi-particles with non-abelian 
exclusion statistics, corresponding to non-abelian braid group 
representations, do not satisfy exclusion statistics of the type 
originally envisaged in [\Ha,\Wuetal].  Recently, however, it was realized
[\GS] that these cases could be incorporated into Haldane's scheme as well
by allowing for pseudo-particles, i.e., particles that do not carry 
any bare mass or energy.  In particular, the non-abelian 
exclusion statistics of $(\wh{\frak{sl}_2})_{k>1}$ spinons (for $k=2,3,4$) 
and generalized fermions in minimal models ${\cal M}_{k+2}$ of CFT (for
$k=1,2,3$) was shown to agree with the pseudo-particle generalization
of Haldane statistics [\GS].

In a parallel development much progress has been made in the last few
years in the analytic calculation of the character formulas directly
from a statistical mechanics approach. These works generally involve
the classification of all the eigenvalues of the transfer matrix and
the computation of their finite-size corrections.  This was first
carried out by the Stony Brook group by solving the Bethe-Ansatz type
equations [\KM], and was followed by the work of [\BPW,\BLZ]  which deals
directly with the functional relations for the eigenvalues.  Typically
these calculations lead to the so-called `fermionic' (or
quasi-particle) type expressions for the characters of the 
representations of the chiral algebra.  One can
again identify the quasi-particles in these fermionic characters, and
they seem to be related to the particle spectrum appearing in certain
integrable perturbations of the underlying CFT [\ABZ,\BMP].

Based on the many known examples (see, e.g., [\BMP--\BSa,\BLSb] 
and references therein), McCoy et al.\ 
(see, in particular, [\BM])
conjectured that all CFT characters can be written in the so-called
`Universal Chiral Partition Function' (UCPF) form, which can be
interpreted as the grand partition function for a system of chiral
particles with fugacities, and whose single particle momenta satisfy
certain fermionic counting rules. Actually, it was noted a few years
earlier by the Stony Brook group [\KMM] that such counting rules are
very similiar to Haldane's exclusion statistics.  The relation of
exclusion statistics to models solvable by the Thermodynamic Bethe
Ansatz (TBA) was also noticed by Bernard
and Wu [\BW].  
Thus it became natural to conjecture that the
quasi-particles underlying the UCPF satisfy Haldane exclusion
statistics with a statistical interaction matrix $G_{ab}$ given by the
bilinear form matrix entering the UCPF.  This was successfully
demonstrated in a number of cases corresponding to abelian exclusion
statistics [\Hik,\ES,\Gai,\BM,\BSc], but it was realized [\BSc], 
and confirmed for
$(\wh{\frak{sl}_2})_{k>1}$ spinons and generalized fermions in [\GS],
that the most general form of the UCPF involves quasi-particles with
non-abelian exclusion statistics.

In this paper we will further elaborate on exclusion statistics in
CFT, and the connection with the UCPF.  In particular, in Sections 2
and 3, we will show that both lead to the same effective central
charge.  Furthermore, in Section 4, we will argue that, in general, to
write characters of WZW models in UCPF form one needs to introduce,
besides the pseudo-particles, yet another kind of quasi-particles.
These particles are composites of the basic quasi-particles and
account for the null-states in the quasi-particle Fock spaces [\BH].
We will refer to these as null-particles.  They can contribute to the
UCPF either with strictly positive or with alternating signs.  We will
incorporate these null-particles in Haldane's scheme from the outset.
In Section 5 we discuss various examples, corresponding to both
abelian and non-abelian exclusion statistics.  We compare the exclusion 
statistics defined by the interaction matrix $G_{ab}$ in the UCPF with 
the results obtained from the recursion relation approach and find 
complete agreement in all cases.  We end with some conclusions.

\newsec{Exclusion statistics with pseudo- and null-particles}

Exclusion statistics, as introduced by Haldane [\Ha] (and generalized 
to the multi-component case in [\FK]%
\foot{For a different approach to
exclusion statistics with internal degrees of freedom, see [\BFa].}),  
is based on the 
idea that the number of accessible states $d_{(a,k)}$ 
for a particle of species
$a$ and momentum $k$
depends on the occupation number $N_{(a,k)}$ of all other particles
though a statistical interaction matrix $g_{(a,k)(b,k')}$ by
\eqn\eqEXbba{
\De d_{(a,k)} \eql - \sum_{(b,k')}\ g_{(a,k)(b,k')}\, \De N_{(b,k')} \,.
}
It follows that the total number of states $W(\{N_{(a,k)}\})$, at fixed 
occupation numbers $\{N_{(a,k)}\}$ is given by
\eqn\eqEXbbb{
W(\{N_{(a,k)}\}) \eql 
\prod_{(a,k)}
\bin{D^0_{(a,k)} + N_{(a,k)} -1 - \sum_{(b,k')}
g_{(a,k)(b,k')} N_{(b,k')} }{N_{(a,k)}}\,,
}
where $D^0_{(a,k)}$ is the total number of states available to
particles of species $a$ with momentum $k$ when there are no particles 
in the system. Thus a gas of particles satisfying the above exclusion 
statistics would have a grand canonical partition function given by 
\eqn\eqEXbaa{
Z \eql \sum_{\{ N_{(a,k)} \} } 
\left( \prod_{(a,k)} (\ta_a)^{N_{(a,k)}} \right) \,
W(\{N_{(a,k)}\}) \exp\left( \sum_{(a,k)}
N_{(a,k)}(\mu_a-
\ep^0_a(k))/k_BT \right)\,,
}
where $\ep^0_a(k)$ and $\mu_a$ are, respectively, 
the bare energy and chemical 
potential of the particle of species $a$.
In the sequel we shall specialize to the case of a one-dimensional
ideal gas where the particle interaction is localized in momentum space
and encoded in a finite matrix $G_{ab}$,  
i.e.,
\eqn\eqLCgG{
g_{(a,k)(b,k')} \eql \de_{k,k'} G_{ab}\,.
} 
We have 
allowed for particles that contribute to the partition function \eqEXbaa\ 
with alternating signs, i.e.\ $\ta_a=-1$, as opposed to the `real
particles' with $\ta_a=1$.  We will see that they occur naturally in
the quasi-particle description of certain conformal field theories.%
\foot{Of course, the alternating sign can be absorbed in the 
exponent by adding an imaginary part to the chemical potential.} We
also partition the full set of particle species
$\cS$ into a set of `physical
particles' $\cSph$, and a set of `pseudo-particles' $\cSps$. The
pseudo-particles do not carry any bare mass or energy (i.e.
$\ep^0_a(k)=0$), but have the unique role of exchanging internal degrees
of freedom (color) between the physical particles.  In
TBA literature they arise in models with non-diagonal
scattering (see, for example, [\YY--\BR]).
Pseudo-particles were recently introduced in Haldane's framework 
in [\GS].  It also seems that they have been anticipated in [\FK] where 
the case of one physical particle and several pseudo-particles was
referred to as a hierarchical basis.  

In the thermodynamic limit where the system size $M\to\infty$ with
finite fixed temperature $T>0$, a saddle point approach to the
partition function \eqEXbaa\ yields the following most probable
1-particle distribution function [\Wuetal]
\eqn\eqEXbab{
n_a(k) \eql z_a {\p\over \p z_a} \log \la_a(z) _{\big| z_b=
\ta_b e^{\be(\mu_b - \ep^0_b(k))}} \,,
}
where $\la_a$ is determined by
\eqn\eqEXbac{
\left({ \la_a - 1\over \la_a} \right) 
\prod_{b} \la_b^{G_{ab}} \eql  z_a \,,
}
and $z_a = \ta_a\exp( \be(\mu_a - \ep^0_a(k)))$.  {}From a TBA point
of view, $\la_a=1+\exp(-\be\ep_a)$ where $\ep_a$ is the dressed energy
for species $a$.  One could proceed further, generalizing the computation
in [\BM], by using $n_a$ to calculate
the internal energy per unit length.  Alternatively, we could work with the
expression for the free energy obtained in [\Wuetal,\FK] 
\eqn\eqLCF{
F \eql -k_B T \sum_{(a,k)}\ D^0_{(a,k)}\, \log \la_a\,.
}
For a gas with linear dispersion relation, i.e.,
\eqn\eqLCdr{
(\ep^0_a(k),\,  D^0_{(a,k)}) \eql \cases{
(v|k|,\,  M \Delta k/ 2\pi) & for $a\in\cSph\,,$ \cr
(0,0) & for $a\in\cSps\,,$ \cr}
}
where we assumed the speed of sound $v$ is independent of the species, 
we obtain   
\eqn\eqLCFb{
F \eql
-{k_B^2 T^2 M\over v} \sum_{a\in\cSph} \int_{0}^{y_a} {dz_a\over{z_a}}\ 
\log\la_a(z) \,,
}
where $y_a=\ta_a e^{\be \mu_a}$ is the fugacity of species $a$.
Thus the specific heat per unit length (at constant fugacity) for a
one-dimensional ideal gas with exclusion statistics \eqEXbba\ is
\eqn\eqLCC{
C \eql {\pi^2 k_B^2 T\over3v}\, \tilde{c}\,,
}
with
\eqn\eqLCc{
{\pi^2 \over6}\tilde{c} \eql \sum_{a\in\cSph} \int_{0}^{y_a} {dz_a\over{z_a}}
\ \log\la_a(z) \,.
}
We have written the specific heat in a form where 
$\tilde{c}$ admits an interpretation
as the effective central charge for systems with conformal symmetry.

The integral \eqLCc\ may be evaluated along the lines of [\BSc]
(similar computations are of course well-known from related 
TBA equations, cf.\ [\BR,\AlBZb,\AlBZa,\KlM,\BFa]) and
leads to\foot{Here we have assumed that $G_{ab}$ is symmetric.}
\eqn\eqEXbe{
\left( {\pi^2 \over6} \right) \tilde{c}(y)
\eql \sum_a \left( L(\xi_a) - L(\et_a) -
\txt{1\over2} \log y_a \log \left({1-\xi_a\over 1-\et_a}\right)  \right)\,,
}
where $(\xi_a, \et_a)$ are solutions to the equations
\eqn\eqEXbf{
\xi_a \eql y_a \prod_b (1-\xi_b)^{G_{ab}}\,,\qquad
\et_a \eql y_a \si_a \prod_b (1-\et_b)^{G_{ab}}\,,
}
where $\si_a=0$ ($\si_a=1$) for $a\in\cSph$ ($a\in\cSps$),
and $L(x)$ is Rogers' dilogarithm defined by
\eqn\eqEXbg{
L(x) \eql - \txt{1\over2} \int_{{\cal C}_{0,x}} \, dz \left(
  {\log z \over 1-z} + {\log (1-z)\over z} \right) \,,
}
where $\log z$ (for $z\neq0$) signifies the logarithm in the branch
$-\pi<{\rm arg}\, z\leq \pi$ and ${\cal C}_{0,x}$ is a contour in $\CC$ from
$0$ to $x$ which does not go across the branch cuts of $\log z$ and
$\log(1-z)$. 
Thus, in contrast to the case with no pseudo-particles [\BSc],
the presence of the pseudo-particles induces subtraction terms
in the effective central charge \eqEXbe.

\newsec{Exclusion statistics in conformal field theory} 

Suppose we have a two-dimensional conformal field theory with
chiral algebra $\cA$, a set of $\cA$-modules $V_i$, labeled by
some index set $i\in\cI$, and intertwiners (CVOs)
\eqn\eqEXbbd{
\phi^a\left( \matrix{ j \cr i'\ i\cr} \right) (z) \eql
\sum_{n\in\ZZ}\ \phi^a\left( \matrix{ j \cr i'\ i\cr} 
\right)_{-n-(\De_{i'} - \De_{i})} \ 
z^{n + (\De_{i'} - \De_{i} -\De_j)}\,,
}
where $\De_i$ denotes the conformal dimension of $V_i$ and
$a=1,\ldots,{\rm dim}\,V_j$.
The number of intertwiners $i\times j \to i'$ is given by the 
fusion rules $N_{ij}{}^{i'}$.
In a quasi-particle approach to conformal field theory the (chiral) 
spectrum is
constructed by repeated application of the modes of a preferred set of 
CVOs on the vacuum $|0\rangle$, i.e.,
a set of quasi-particle states of type 
\eqn\eqEXbbf{
\phi^{a_N}\left( \matrix{ j_N \cr i_{N}\ i_{N-1} \cr} \right)_{-n_N-\De(N)}
\ldots 
\phi^{a_2}\left( \matrix{ j_2 \cr i_{2}\ i_{1} \cr} \right)_{-n_2-\De(2)}
\phi^{a_1}\left( \matrix{ j_1 \cr i_{1}\ 0 \cr} \right)_{-n_1-\De(1)}
|0\rangle\,,
}
where $\De(k) = \De_{i_k} - \De_{i_{k-1}}$,
constitute a basis of the $\cA$-module $V_i$.  This basis is overcomplete
unless we put restrictions on the mode sequences $(n_1,\ldots,n_N)$.
These restrictions are obtained both from the 
braiding and fusion relations satisfied by 
the intertwiners \eqEXbbd\ 
as well as by possible null-states in the Fock space 
of intertwiners and may depend on the fusion path $(0,i_1,i_2,\ldots,i_N)$.

It has been observed in a variety of approaches -- TBA approaches,
integrable spin chains and also in the context of conformal field
theory -- that the degrees of freedom contained in \eqEXbbf\ can be
separated into physical excitations and pseudo-particle excitations.
Loosely speaking,
the physical excitations correspond to excitations over some reference
fusion path, while the pseudo-particles create `excited fusion paths'.
While this separation might, strictly speaking, not hold in 
the conformal field
theory, it may hold in some crystal limit (cf.\ [\NYa]) which is
sufficient as far as the discussion of the partition function is
concerned.  Thus, if the quasi-particles in a conformal field theory
are described by Haldane exclusion statistics, we need to distinguish
two kinds of particles: `pseudo-particles' with 
vanishing bare energy and `physical
particles' with an infinite range of energy levels separated by
integers, i.e., we have the dispersion relation \eqLCdr.

Let us consider the partition function ${\rm ch}_i(y;q)$, i.e., 
the character of the $\cA$-module $V_i$. 
If we assume that the quasi-particle interaction is purely statistical
according to \eqEXbba, and that
$\la_a$ of \eqEXbac\ can be interpreted as the single quasi-particle 
grand partition function, the character will have the following 
approximate form in the thermodynamic limit%
\foot{The modular parameter $q$ is related to the 
quantum spin chain quantities by $q = \exp({-2\pi{v}/Mk_BT})$, thus
$M\to\infty$ (at fixed $T>0$) corresponds to $q\to1^-$.}
\eqn\eqEXbad{
{\rm ch}_i(y;q) ~\sim~ \prod_{a\in\cSps} \la_a(y_a q^{\De_a})
\prod_{a\in\cSph} \prod_{l\geq0} \la_a(y_aq^{\De_a + l})\,. 
}
Of course the expression \eqEXbad\ is not meant to be exact but only
valid, in general, in the thermodynamic limit.  We have chosen
to write it in a discrete form, rather than in an integral form like 
\eqLCFb, to emphasize the discrete energy
spectrum of the CFT.  Also, the product over $a$ and $l$ will be
subject to restrictions depending on the sector $i$.  Cases where the
CFT characters (or sums thereof) do admit exact factorizations of the
form \eqEXbad\ have recently been studied in [\BFb].

The character ${\rm ch}_i(y;q)$ may be expanded as a power series in $q$
\eqn\eqEXba{
{\rm ch}_i(y;q) \eql \sum_{N\geq0} a_N(y)\, q^N\,, } 

It is well-known, of course, that modular transformations relate the
asymptotic behaviour of $a_N(y)$, for $N\gg0$, to the specific heat
\eqLCC.  For definiteness, let us see how this works out using \eqEXbad.
Asymptotically, we may approximate
\eqn\eqEXbc{ \eqalign{
a_N(y) \eql & {1\over 2\pi i}\oint {dq\over q^{N+1}} \ 
{\rm ch}_i(y;q) \eql {1\over 2\pi i}
\oint dq \ \exp\left( -(N+1)\log q + \log {\rm ch}_i(y;q)\right) \cr
  ~\sim~ &  \oint dq \exp\left(-(N+1)\log q  
- (\log q)^{-1} \sum_{a\in\cSph} \int_0^{y_a} {dz_a\over z_a} 
\log \la_a(z)\right)\,.\cr}
}
In the last step, we have omitted all terms that do not contribute
to the leading $N$ behaviour of $a_N(y)$.
The integral can be evaluated using a saddle point approximation, 
and we find
\eqn\eqEXbb{
\log a_N(y) ~\sim~ 2\pi \sqrt{ {\ceff(y) N\over 6}}\,,
}
with 
\eqn\eqEXbd{
{\pi^2 \over 6}\, \ceff(y) \eql \sum_{a\in\cSph} \int_0^{y_a} {dz_a\over z_a}
\log \la_a(z)\,.  
}
{}From \eqEXbb\ we 
can identify $\ceff(y)$ with the effective central charge of the 
partition function \eqEXbad\ as shown in, e.g., [\NRT,\DKKMM].
Note that \eqLCc\ and \eqEXbd\ indeed imply that 
we can also identify $\ceff(y)$ with the $\tilde{c}$ computed in the
previous section.

For future use, note that if all the chemical potentials $\mu_a$, for
$a\in\cSph$, are 
given in terms of a single chemical potential $\mu$ as $\mu_a = \ell_a \mu$,
then we may write \eqEXbd\ as 
\eqn\eqEXbda{
{\pi^2 \over 6}\, \ceff(y) \eql \int_0^{y} {dz\over z}
\log \la_{\rm tot}(z)\,,
}
where 
\eqn\eqEXbdb{
\la_{\rm tot}(z) \eql \prod_{a\in\cSph} \ \la_a(z)^{\ell_a}\,.
}

The central charge $c=\ceff + \De_{\rm min}$ of the conformal field 
theory follows from \eqEXbe\ and \eqEXbf\ in the limit 
of vanishing chemical potentials, i.e.\ $y_a=\ta_a$,
where we need to take those solutions to \eqEXbf\ that satisfy 
$0\leq \xi_a,\et_a \leq1$ for $\ta_a=1$ and $-1\leq\xi_a,\et_a\leq0$
for $\ta_a=-1$.  For the latter case, the imaginary part of the corresponding
dilogarithm is precisely canceled by the logarithm term in \eqEXbe.  Indeed,
for $x<0$,
\eqn\eqEXbh{
L(x) - {\pi i\over2} \log(1-x) \eql L\left( {1\over 1-x} \right) - L(1) 
\eql -L\left( {-x\over 1-x} \right)\,.
}

An interesting development over the last few years has been 
the derivation of quasi-particle type character formulas for the
modules of chiral algebras (see, e.g., [\BMP--\BSa,\BLSb]).
This work has led to the 
conjecture that all conformal field theory characters can be written
in the so-called `Universal Chiral Partition Function' (UCPF) form 
(see, in particular, [\BM])
\eqn\eqEXca{
{\rm ch}_i(y;q) \eql \sum_{m_1,\ldots, m_n\geq0 \atop {\rm restrictions} }
 \left( \prod_{a} y_a^{m_a} \right) 
 q^{ {1\over2} \bm\cdot \bG \cdot \bm - {1\over2} 
\bA \cdot \bm}\prod_a \qbin{ ((\id-\bG)\cdot \bm + {\bu\over2})_a}{m_a} \,,
}
where $\bG$ is an $n\times n$ matrix and $\bA$ and $\bu$ are certain 
$n$-vectors.  Both $\bA$ and $\bu$ as well as the restrictions on the
summations over the quasi-particle numbers $m_a$ will in general depend
on the sector $i$, while $\bG$ will be independent of $i$.
Furthermore, we have defined
\eqn\eqEXta{
\qbin{m}{n} \eql {(q)_m\over (q)_n (q)_{m-n} }\,,\qquad\quad
(q)_n \eql \prod_{k=1}^n\, (1-q^k) \,.
}
It has been conjectured by various groups (see, in particular,
[\BM]) that the quasi-particles underlying \eqEXca\ satisfy Haldane 
exclusion statistics with statistical interaction matrix
given by the {\it same} matrix $\bG$ as the one entering \eqEXca.
We will refer to this conjecture as the `UCPF-exclusion statistics' 
conjecture.

A convincing piece of evidence in support of this conjecture is that
the asymptotics of the character \eqEXca\ (in the thermodynamic limit
$q\to1^{-}$) is given by exactly the same formulas \eqEXbe\ and
\eqEXbf\ where $\si_a=0$ for $u_a=\infty$ (physical particles), while
$\si_a=1$ for $u_a<\infty$ (pseudo-particles).  The asymptotic form of
the character \eqEXca\ for $y_a=1$ was derived in [\DKKMM] (see also
[\RS,\NRT,\Kir]).  The present result is a straightforward generalization
of these derivations.

To prove the conjecture beyond the comparison of thermodynamics
requires an exact computation of the partition function starting from
first principles, i.e., eqn.\ \eqEXbba, as has been done for $g$-ons
[\Hik].  Alternatively, it has been argued that the analytic
continuation of \eqEXbe\ to the covering space of $\CC\backslash
\{0,1\}$ contains information about {\it all} the excitations in the
spectrum.  This idea has been successfully applied to some minimal
models of conformal field theory [\BPW], (generalized) parafermions
[\KN], $(\widehat{\frak{sl}_n})_{k=1}$ WZW models [\Sub] and
$(\widehat{\frak{sl}_2})_{k\geq1}$ WZW models [\Sua] and might be put
on a more rigorous footing.

Exclusion statistics in conformal field theory can be studied 
by a method based on recursion relations for truncated characters [\Sca].
Truncated characters $P^{(i)}_L(y;q)$ are defined by taking the partition 
function
of those states \eqEXbbf\ where all the modes satisfy $n_i+\De(i)\leq L$.
Thus, for large $L$, we will have (cf.\ \eqEXbad)
\eqn\eqEXbi{
P_L^{(i)}(y;q) ~\sim~ \prod_{a\in\cSps} \la_a(y_a)
\prod_{a\in\cSph}\ \prod_{0\leq l\leq L} \la_a(y_aq^{l})\,,
}
where the products are subject to certain restrictions depending on
the sector $a$. Thus
\eqn\eqEXbj{
P_{L+1}^{(i)}(y;q) /P_L^{(i)}(y;q) ~\sim~\prod_{a\in\cSph}
  \la_a(y_a q^{L}) \,, \qquad {\rm as\ }{L\to \infty} \,.
}
More generally, if some of the physical particles $a\in\cSph$ 
are composites of
$\ell_a$ more fundamental particles, then their modes will be cut
off at $\ell_a L$ and we find 
\eqn\eqEXbk{
P_{L+1}^{(i)}(y;q) /P_L^{(i)}(y;q) ~\sim~ \prod_{a\in\cSph} \la_a(y_a
  q^{L})^{\ell_a} \eql \la_{\rm tot}(y_aq^L)\,, 
} 
where $\la_{\rm tot}$ is defined in \eqEXbdb.  Therefore, from
recursion relations for the truncated characters $P_L^{(i)}(y;q)$ in
the large $L$ limit, one derives algebraic equations for the total
1-particle partitions functions $\la_{\rm tot}(y)$, which can be
compared to the $\la_{\rm tot}(y)$ arising from a solution of
\eqEXbac, with a statistical interaction matrix as suggested by the
UCPF formula \eqEXca.  This procedure was carried out, and agreement
was found, in several cases including $g$-ons (the one-component case
of \eqEXca) [\Hik,\BM] and several multi-component cases
[\Sca,\ES,\Gai,\BSc].  All these cases involve only physical particles
($u_a=\infty$), i.e., correspond to intertwiners with a unique fusion
path, and the corresponding statistics was therefore called `abelian
exclusion statistics'.  {}From \eqEXbac\ it is clear that the absence
of pseudo-particles always leads to small $x$ expansions of the form
$\la_a(x) = 1 + x_a + \cO(x^2)$.  In [\Scb,\BSc] it was observed,
however, that generally we have expansions of the form $\la_a(x) = 1 +
\al_a x_a + \cO(x^2)$, where $\al_a$ is the largest eigenvalue of the
fusion matrix of the quasi-particle $a$ [\BSc].  The exclusion
statistics corresponding to the more general case, $\al_a\neq1$, was
named `non-abelian exclusion statistics'.  It was recently recognized
that non-abelian exclusion statistics can be accounted for in the
Haldane approach by incorporating pseudo-particles [\GS]. The
UCPF-exclusion statistics conjecture was subsequently confirmed in
various non-abelian cases, namely $(\wh{\frak{sl}_2})_{k>1}$
WZW-models and generalized fermions in minimal models of CFT [\GS].

The main purpose of the remainder of this paper is to verify the
UCPF-exclusion statistics conjecture in various other, rather
non-trivial, WZW-examples (both abelian and non-abelian) for which
UCPF characters have been found recently [\BH].  In all examples we
find complete agreement, thus supporting the conjecture above.  To
find agreement, however, we have had to slightly extend the definition
of the UCPF form to account for certain null-particles, as remarked
before.

\newsec{Towards the UCPF for WZW models}

The following questions naturally arise.  Given a conformal field 
theory, how does one identify a set of quasi-particles in terms of which
the characters take the UCPF form, and how is the statistical interaction
matrix $\bG$ determined in terms of the conformal field theory data
(chiral algebra, modules, fusion rules, conformal dimensions, etc.)?
In this section we will give a partial answer for WZW models. 

Thus far, only isolated cases of UCPF characters for affine Lie algebras
were known.  These included $(\wh{\frak{sl}_2})_{k=1}$ [\BPS,\BLSa],
$(\wh{\frak{sl}_2})_{k>1}$ [\BLSb] and $(\wh{\frak{sl}_n})_{k=1}$ 
[\BSa,\BSd].%
\foot{In principle one can get quasi-particle affine Lie algebra characters 
by taking limits of the $\cW$-algebra minimal model characters (i.e.,
coset characters) of, e.g., [\DKKMM].  These will however involve an infinite
number of pseudo-particles and are not of the type considered here.}
Recently an algorithm was given which, in principle, can be used to
obtain affine Lie algebra characters in UCPF form for any affine Lie
algebra $\whg$ and at arbitrary level [\BH].  
Here we will briefly explain the idea, we
refer to [\BH] for the technical details.  In the next section we
discuss some examples, examine the exclusion statistics, and make a
comparison to the results of [\BSc].

Let $\bfg$ be a simple finite dimensional Lie algebra of rank $\ell$,
let $\{\La_i\}_{i=1}^\ell$ be the set of fundamental weights and $L(\La_i)$
the corresponding finite dimensional irreducible representations.
As our set of quasi-particles we take the intertwiners%
\foot{While in some level-$1$ cases it is possible to give the character 
in a UCPF form using less quasi-particles than the ones discussed
here, these formulas do not seem to generalize to level $k>1$ just
by the inclusion of additional pseudo-particles.  We refer to Section
5.6 for an example.}
\eqn\eqEXcaa{
\phi^a\left( \matrix{ \La_i \cr \La' \ \La \cr}\right)(z)\,,
\qquad a=1,\ldots, {\rm dim}\ L(\La_i)\,,
}
corresponding to all fundamental representations $L(\La_i)$ and between all
possible $\whg$ modules (given by $\La$ and $\La'$) at level-$k$, as 
determined by the fusion rules $N_{\La\La_i}{}^{\La'}$.
For example, for $\wh{\frak{sl}_3}$ we take intertwiners transforming 
in both the $\bf3$ as well as the $\bar{\bf3}$ 
representation of $\frak{sl}_3$.
Next, we need to decouple the pseudo-particle excitations, representing
the excited fusion paths with respect to a reference fusion path, from
the physical excitations.  The sums over pseudo-particle excitations 
are well known from RSOS and spin chain
models and yield, up to a factor, level-$k$ restricted modified
Hall-Littlewood polynomials $M^{(k)}_{\la\mu}(y;q)$ 
(or Kostka-Foulkes polynomials in the case of $\frak{sl}_n$)
for which, in some cases, UCPF expressions are known (see, e.g.,
[\BMS,\DKKMM,\Kir,\HKKOTY] for $y=1$).  
On the other hand, the physical excitations would yield 
a (Fock space) contribution to the character given by
\eqn\eqEXcb{
\sum_{\sum_a m^{(i)}_a = M_i} \ \left( \prod_i\, y_i^{M_i} \right)\,
{q^{ {1\over2} \bM \cdot \bB \cdot \bM }
\over \prod_i \prod_{a} \qn{m^{(i)}_a} } \,,
} 
where $M_i$ is the total number of intertwiners in $L(\La_i)$ and
we have specialized to the case where all particles in $L(\La_i)$ have
the same fugacity $y_i$.  The bilinear form matrix $\bB$ is given,
in the case of simply laced Lie algebras, by the inverse Cartan 
matrix $A_{ij}^{-1} =(\La_i,\La_j)$ of $\bfg$ and arises from the 
mutual locality of the basic vertex operators $\exp(i\La_i\!\cdot\! \ph)$.
For non-simply laced Lie algebra these vertex operators have to be corrected
by fermionic operators to account for the difference between $\half
|\La_i|^2$ and the conformal dimension $\De_i = (\La_i,\La_i+2\rh)/
({\rm h}^\vee +1)$ of $V_i$ at level $k=1$ -- the corresponding 
$\bB$ easily follows from [\GNOS].
Combining these two ingredients gives the
required result for, e.g., $(\wh{\frak{sl}_2})_{k\geq1}$ (see Section
5.1).  Unfortunately, in general this is not the whole story as we
have not yet incorporated the possible null-states in the physical
quasi-particle Fock space.  It turns out, as discussed in [\BH],
that these can be accounted for in UCPF form by interpreting the Fock
space modulo null-states as the coordinate ring of an affinized
projective variety and applying standard techniques from algebraic
geometry.  Besides deforming the exponent ${1\over2}\bM\!\cdot\! \bB
\!\cdot\! \bM$ in
\eqEXcb\ by a term ${1\over2} \bm\!\cdot\! {\bf Q}\!\cdot\! \bm$, this will
in general involve the addition of null quasi-particles,
corresponding to certain composites of the basic physical
quasi-particles (hence their chemical potentials are fixed in terms of
those of their constituents).  

The final answer for the UCPF is then of the form
\eqn\eqEXlc{
{\rm ch}_\la(y;q) \eql \sum_{\mu=M_1\La_1+\ldots+M_\ell\La_\ell}
{1\over \prod_{i=1}^\ell \qn{M_i} } \ 
M_{\la\mu}^{(k)}(y;q) M_\mu(y;q)\,,
}
where 
\eqn\eqEXld{
M_\mu(y;q) \eql \left( \prod_{i=1}^\ell \qn{M_i} \right) \sum_{\bm}\
\left( \prod_a y_a^{m_a} \right) 
{ q^{ {1\over2} \bm \cdot {\bf Q} \cdot \bm } \over \prod_a \qn{m_a} }\,.
}
The factor containing $\bB$ in \eqEXcb\ has been absorbed in
$M_{\la\mu}^{(k)}(q)$. 
Indeed, it has been conjectured (and proven in some cases)
[\NYc,\Ya,\HKKOTY] that the 
affine Lie algebra characters ${\rm ch}_\la(y;q)$ are indeed of 
the form \eqEXlc, where $M_{\la\mu}^{(k)}(y;q)$ and $M_\mu(y;q)$ are,
respectively, the level-$k$ restricted and unrestricted Hall-Littlewood
polynomials.

The procedure leading to \eqEXld\ is not unique, however,
and various equivalent UCPFs with different null quasi-particle
contents may be given (see the examples in Section 5).%
\foot{An interesting example in the context of minimal models of 
conformal field theory was recently discussed in [\BMP].}

The equality between the characters with the various null-state 
subtractions is based on the successive application of the 
following two identities (see, e.g., [\BSd] for an elementary proof)
\eqn\eqEXch{
{ 1\over \qn{M} \qn{N} } \eql \sum_{m\geq0} {q^{(M-m)(N-m)} \over 
\qn{m}\qn{M-m}\qn{N-m} } \,,
}
and
\eqn\eqEXcg{
{ q^{MN}\over \qn{M} \qn{N} } \eql \sum_{m\geq0} (-1)^m {q^{ {1\over2}m(m-1)}
\over \qn{m}  \qn{M-m}\qn{N-m} } \,,
}
which are, in a sense, `inverses' of each other.
Both identities are specializations of the $q$-Chu-Vandermonde identity
for the basic hypergeometric series $_2\ph_1$ (see, e.g., [\GR]).%
\foot{We are grateful to Ole Warnaar for pointing this out to us.}
Both are intimately related to, and in fact constitute a proof of,
the five-term identity for Rogers' dilogarithm 
\eqn\eqEXci{
L(x) + L(y) \eql L(xy) + L\left( {x(1-y)\over 1-xy} \right) +
L\left( {y(1-x)\over 1-xy} \right) \,.
}
Indeed, by comparing $\ceff(y)$ for the asymptotics
of two characters related by a single application of either \eqEXch\ 
or \eqEXcg\ one discovers \eqEXci.
Denoting the solutions of \eqEXbf\ for the corresponding 
variables on the left hand side of \eqEXch\ 
and \eqEXcg\ by $(\xi_1,\xi_2)$ and on the right hand side by 
$(\xi_1',\xi_2',\xi_3')$ we find that for \eqEXch\ they are related 
by 
\eqn\eqEXcj{
\xi_1' \eql {\xi_1(1-\xi_2) \over 1-\xi_1\xi_2}\,,\qquad
\xi_2' \eql {\xi_2(1-\xi_1) \over 1-\xi_1\xi_2}\,,\qquad
\xi_3' \eql \xi_1\xi_2 \,,
}
while for \eqEXcg\ we find the inverse relations
\eqn\eqEXck{
\xi_1 \eql {\xi_1'(1-\xi_2') \over 1-\xi_1'\xi_2'}\,,\qquad
\xi_2 \eql {\xi_2'(1-\xi_1') \over 1-\xi_1'\xi_2'}\,,\qquad
\xi_3' \eql - {\xi_1'\xi_2' \over 1-\xi_1'\xi_2'}\,.
}
With the use of \eqEXbe\ and \eqEXbh\
both lead to \eqEXci.
The fact that the characters of the various null-particle formulations 
always seem to be related by either \eqEXch\ 
or \eqEXcg\ can be taken as further evidence for the conjecture 
that, loosely speaking, 
dilogarithm identities are always accessible by means of the 
five-term identity (`Goncharov's conjecture', see [\Kir]).

Having obtained the WZW characters in a UCPF form we can read off the
statistical interaction matrix $\bG$ and verify whether the alleged
exclusion statistics defined by \eqEXbac\ indeed agrees with the
exclusion statistics derived by the recursion method in [\BSc].  We
will carry out this procedure in several non-trivial examples 
(Section 5) and find agreement in all cases.

In [\BSc] one of the authors and K.~Schoutens
conjectured that the recursion relations for the 
truncated characters $P_L^{(i)}(q)$ of the level-$1$
WZW models were, in all cases,
solved (upto an overall $q$-factor )
by modified Hall-Littlewood polynomials $M_{\la(L,i)}(q)$,
with argument $q\to q^{-1}$, and $\la$ a function of both $i$ and $L$.
Thus, these modified Hall-Littlewood polynomials have to approach the 
WZW characters ${\rm ch}_i(q)$ in the limit $L\to\infty$.  
This observation leads to the conclusion
that, asymptotically, $M_{\la(L,i)}(q)$ is given by an expression like
\eqEXbi.  Extrapolating this reasoning a bit further, using the fact
that $M_{\la}(q)$, for $\la=M_1\La_1+\ldots+M_\ell\La_\ell$,
is precisely the TBA-limit of the partition function of an integrable
spin chain with $M_i$ spins in the minimal $U_q(\whg)$ affinization $W_i$ of
the fundamental representation $L(\La_i)$
whose elementary excitations are precisely the 
quasi-particles \eqEXcaa\ [\KR], we are led to the conjecture that,
asymptotically,
\eqn\eqEXqa{
M_\la(q) ~\sim~ q^{N(\la)} \prod_{i=1}^\ell \prod_{l_i=1}^{M_i} 
 \ze_i(q^{-l_i})\,,
}
where $N(\la)$ is such that $M_\la(q)={\rm const} +\cO(q)$ and the $\ze_i$ 
are expressed in terms of the $\la_a$ according to the fusion rules, i.e.,
according to which composites of the quasi-particles
$\phi^a\left( \matrix{ \La_i \cr \La' \ \La \cr}\right)(z)$ make up the 
sectors $j$.  For $\frak{sl}_n$ we would have, more explicitly,
\eqn\eqEXqb{
\ze_i(x) \eql \prod_j \left( \prod_{a} \wt{\la}_a^{(j)}(x) 
\right)^{B_{ij}} \,,
}
where $B_{ij}$ is the inverse Cartan matrix of $\frak{sl}_n$ and 
$\wt{\la}_a^{(j)}$ are the solutions ${\la}_a^{(j)}$ 
to \eqEXbac\ for the physical particles corresponding to \eqEXcaa\ dressed 
with the $\la$'s for the composite null-particles which contain that
particle.  Explicit formulas for $\ze_i$ will be given in the examples
of Section 5.  Note however that the modified Hall-Littlewood 
polynomial is a $q$-deformation of the character of the tensor product 
of $M_i$-fold copies of $W_i$ .  
Thus, a consistency check on the assertion \eqEXqa\ is that
\eqn\eqEXqc{
\ze_i(1) \eql {\rm dim}\,W_i\,.
}
We will verify this in the examples.  In fact, the analogous statement 
seems to be true in higher level cases as well as suggested by the 
$(\wh{\frak{sl}_2})_k$ example in Section 5.1.

The fact that limits of modified Hall-Littlewood (or Kostka-Foulkes) 
polynomials lead to WZW characters for $(\wh{\frak{sl}_n})_k$ 
was first conjectured in [\Kir] and subsequently proven, in
special cases, in [\NYb,\HKKOTY].  

\newsec{Examples}

\subsec{$\wh{\frak{sl}_2}$, level-$k$}

The (non-abelian) exclusion statistics for the case
$(\wh{\frak{sl}_2})_k$ has been extensively discussed in
[\FKY,\FrS,\BSc,\GS], where it was shown, among other things, that 
the solutions to the equation \eqEXbac\ indeed agree with the expressions
obtained from the recursion approach (at least for small $k$).  
Here we suffice by making a few additional remarks.

At level-$k$ there are $k+1$ integrable highest
weight modules of $\wh{\frak{sl}_2}$ labeled by (twice) the $\frak{sl}_2$
spin, $2j=0,1,\ldots,k+1$.
The character of $(\wh{\frak{sl}_2})_{k\geq1}$ can be written as 
[\BLSb,\NYa,\ANOT]
\eqn\eqEXzg{ \eqalign{
{\rm ch}_j(z;q) \eql &
\sum_{m_+,m_-\geq 0 \atop m_++m_-=m_1} \sum_{m_2,\ldots,m_k\geq0 \atop
{\rm restrictions}} q^{-{1\over4}m_1^2+ {1\over4} \bm\cdot  \bA_k  \cdot \bm}
\prod_{a=2}^k \qbin{((\id - {1\over2} \bA_k)\cdot \bm + {\bu_j\over2})_a}{m_a}
\cr & \times 
{1 \over \qn{m_+}\qn{m_-}} z^{{1\over2} (m_+-m_-)}\,, \cr}
}
where $\bA_k$ is the Cartan matrix of $A_k \cong \frak{sl}_{k+1}$, and
$(\bu_j)_a = \de_{a,2j}$.  The restrictions are such that all 
entries in the $q$-binomials are integers.
This character is obviously of the UCPF form \eqEXca\ with 
\eqn\eqEXza{
\bG \eql \left( \matrix{
\half & \half & \vdots & -\half & & \cr
\half & \half & \vdots & -\half & & \cr
\ldots & \ldots & \ldots & \ldots & \ldots & \ldots \cr
-\half & -\half & \vdots &  & & \cr
 & & \vdots &  & \half\bA_{k-1} & \cr
 & & \vdots & & & \cr} \right) \,,
}
where the entries of $\bG$ correspond to the summation variables
$\{m_+,m_-,m_2,\ldots,m_k\}$ in \eqEXzg.  In particular $u_+=u_-=\infty$
while $u_a<\infty$ for $a=2,\ldots,k$.

We find the following solution to \eqEXbf\ for $y_+=y_-=1$
\eqn\eqEXzb{ \eqalign{
\xi_+ \eql \xi_- \eql 1- {1\over k+1} & \,, \qquad \xi_a \eql 1 -
  \left({1\over k+2-a}\right)^2 \,,\quad a=2,\ldots,k\,,\cr
\et_+ \eql \et_- \eql 0 & \,, \qquad \et_a \eql 1 - \left( 
 {\sin {\pi\over k+2} \over \sin {\pi a\over k+2}} \right)^2 
\,,\quad a=2,\ldots,k\,,\cr}
}
leading to 
\eqn\eqEXzc{ \eqalign{
\left({\pi^2 \over 6} \right)c 
\eql & \sum_a \left( L(\xi_a) - L(\et_a) \right) 
\eql  L(\xi_+) + L(\xi_-) + \sum_{a=2}^k \left( L(1-\et_a) - 
L(1-\xi_a) \right) \cr
\eql & 2L\left( {k\over k+1}\right) - \sum_{a=1}^k L\left( {1\over a^2}\right)
+ \sum_{a=1}^k L\left( \left({\sin {\pi\over k+2} \over 
\sin {\pi a\over k+2}}\right)^2
\right) 
\eql \sum_{a=1}^k L\left( \left( {\sin {\pi\over k+2} \over 
\sin {\pi a\over k+2}} \right)^2 \right) \cr
 \eql & {\pi^2\over6}\left( {3k\over k+2} \right) \,,\cr}
}
as required.  Moreover, as shown in [\Sca] for $k=1$ and 
[\GS] for $k=2,\ldots,4$, the solution to \eqEXbac\ agrees with the one found 
by the recursion method [\FrS,\BSc].
Also note that from \eqEXzb\ it follows that the expression
\eqn\eqEXze{
\ze \eql (\la_+ \la_-)^{1\over2} \,,
}
at $x=1$ is given by $\ze= k+1$, which is consistent with the 
interpretation of the quasi-particles as the excitations of an
$SU(2)$ spin chain with $2S=k$.

\subsec{$\wh{\frak{sl}_3}$, level-$1$}

The affine Lie algebra $\wh{\frak{sl}_3}$, at level-$k=1$, has
three integrable representations corresponding to the singlet
$\bf 1$, the vector ${\bf 3} = L(\La_1)$ and the conjugate vector
$\bar{\bf 3} = L(\La_2)$ of $\frak{sl}_3$.
As discussed in Section 2, for the quasi-particles we take 
the intertwiners $\ph^a(z)$ and $\phi^{\bar a}(z)$ transforming in,
respectively, the $\bf3$ and $\bar{\bf3}$ representations.  
Since at level $k=1$ the
fusion path is unique, there will be no pseudo-particles.  However,
the quasi-particle Fock space will contain null-states as a consequence
of the null-field $\sum_a :\!\ph^a(z)\phi^{\bar a}(z)\!:$.  The most
natural way of incorporating this null-field is by introducing one 
null-particle with $\ta=-1$.  The following character formula for the
integrable highest weight modules of $(\wh{\frak{sl}_3})_{k=1}$ was
found in [\BSa]:
\eqn\eqEXdaa{ 
{\rm ch}_i(y;q) \eql 
 \sum_{M_1,M_2\geq0\atop M_1+2M_2\equiv i\, {\rm mod}\,3} y_1^{M_1}
   y_2^{M_2}
q^{ {1\over2}\bM\cdot \bB\cdot \bM } \ 
\sum_{m_a, m_{\bar a} , m\atop{
\sum m_a + m =M_1\atop \sum m_{\bar a} + m =M_2} } \ 
(-1)^m {q^{ {1\over2}m(m-1) } \over \qn{m} } 
{1\over \prod_a \qn{m_a}  \qn{m_{\bar a}} } \,,
}
where 
\eqn\eqEXdac{
\bB \eql  \left( \matrix{ \twth & \onth \cr \onth &
\twth \cr} \right) \,,
}
is the inverse Cartan matrix of $\frak{sl}_3$.  This is indeed 
of the UCPF form \eqEXca\ with 
\eqn\eqEXdb{
\bG \eql \left( \matrix{ 
\twth & \twth & \twth & \vdots & \onth  & \onth & \onth & \vdots & 1 \cr
\twth & \twth & \twth & \vdots & \onth  & \onth & \onth & \vdots & 1 \cr
\twth  & \twth & \twth & \vdots & \onth  & \onth & \onth & \vdots & 1 \cr
\ldots & \ldots & \ldots & \ldots & \ldots & \ldots & \ldots & 
  \ldots & \ldots \cr
\onth  & \onth & \onth & \vdots & \twth  & \twth & \twth & \vdots & 1  \cr
\onth  & \onth & \onth & \vdots & \twth  & \twth & \twth & \vdots & 1  \cr
\onth  & \onth & \onth & \vdots & \twth  & \twth & \twth & \vdots & 1 \cr
\ldots & \ldots & \ldots & \ldots & \ldots & \ldots & \ldots & 
  \ldots & \ldots \cr
 1&1&1&\vdots  &1&1&1&\vdots & 3\cr} \right) \,.
}
and 
\eqn\eqEXdaca{ \eqalign{
{\bf \ta} & \eql \{ 1,1,1 \,|\,1,1,1 \,|\,-1\} \,,\cr
\bu & \eql \{ \infty,\infty,\infty\,|\,\infty,\infty,\infty\,|\,\infty\}
\,.\cr}
}
As remarked in Section 2, the fugacity of the null particle 
is given in terms of that of its constituents as $-y_1y_2$.
The central charge \eqEXbe\ works out correctly, as $c=2$, with
$\{\xi_a\} = \{ \scr{1\over3},\scr{1\over3},\scr{1\over3}|\scr{1\over3}, 
\scr{1\over3},\scr{1\over3}|-\scr{1\over8} \}$.

To compare the exclusion statistics based on the statistical interaction
matrix \eqEXdb\ with the results of [\Sca,\BSc] we have to solve \eqEXbac\
with 
\eqn\eqEXlb{ \eqalign{
\{ \la_a \} \eql &  \{ \la_1,\la_2,\la_3\,|\,\la_{\bar1},\la_{\bar2},
\la_{\bar3}\,|\,\mu\} \,,\cr
\{ z_a \} \eql & \{ x,x,x|x^2,x^2,x^2| -x^3\} \,. \cr}
}
We find 
\eqn\eqEXda{ \eqalign{
\la & \eqv \la_1 \eql \la_2 \eql \la_3 \eql {1+2x\over 1+x} \eql
1+ {x\over1+x} \,,\cr
\bar\la & \eqv \la_{\bar1} \eql \la_{\bar2} \eql \la_{\bar3} 
\eql {1+x+x^2 \over 1+x} \eql 1+ {x^2\over1+x} \,,\cr
\mu & \eql {(1+x)^3 \over (1+2x)(1+x+x^2) } \eql 1 - {x^3\over 
(1+2x)(1+x+x^2) } \,.\cr}
}
This indeed implies 
\eqn\eqEXdba{
\la_{\rm tot}(x) \eql \left(\prod \la_a \right)
\left(\prod \la_{\bar a} \right)^2 \mu^3 \eql
(1+x+x^2)^3\,,
}
in accordance with the results of [\Sca,\BSc].

Alternatively, we might also incorporate the effect of the null-field 
by slightly changing the statistics of the physical particles $\ph^1$
and $\ph^{\bar1}$.  In the characters this amounts to applying 
\eqEXcg.  This yields
\eqn\eqEXdab{ 
{\rm ch}_i(y;q) \eql 
 \sum_{M_1,M_2\geq0\atop M_1+2M_2\equiv i\, {\rm mod}\,3}
y_1^{M_1}y_2^{M_2}\,q^{ {1\over2}\bM\cdot \bB\cdot \bM } \ 
\sum_{m_a, m_{\bar a}\atop{ 
\sum m_a =M_1,\atop \sum m_{\bar a} =M_2 }} \ 
 {q^{m_1m_{\bar1}} \over \prod_a \qn{m_a}  \qn{m_{\bar a}} }
}
which is of the UCPF form with 
\eqn\eqEXdf{ 
\bG \eql \left( \matrix{ 
\twth & \twth & \twth & \vdots & \scr{4\over3}  & \onth & \onth  \cr
\twth & \twth & \twth & \vdots & \onth  & \onth & \onth  \cr
\twth  & \twth & \twth & \vdots & \onth  & \onth & \onth  \cr
\ldots & \ldots & \ldots & \ldots & \ldots & \ldots & \ldots \cr
\scr{4\over3}  & \onth & \onth & \vdots & \twth  & \twth & \twth  \cr
\onth  & \onth & \onth & \vdots & \twth  & \twth & \twth  \cr
\onth  & \onth & \onth & \vdots & \twth  & \twth & \twth  \cr} \right) \,.
}
The corresponding solutions to \eqEXbac\ are
now given by
\eqn\eqEXde{ \eqalign{
\la'_1 & \eql \la \mu \,,\qquad \la'_2 \eql \la'_3 \eql \la \,,\cr
\la'_{\bar1} & \eql \bar\la \mu \,,\qquad 
\la'_{\bar2} \eql \la'_{\bar3} \eql \bar\la \,.\cr}
}
where $\la, \bar\la$ and $\mu$ are as in \eqEXda.  In other words,
changing the statistics of the physical particles $\ph^1$
and $\ph^{\bar1}$ precisely corresponds to dressing these 
particles by the null-particle in the previous formulation.
Again we find that $\latot = (\prod \la_a') (\prod \la_{\bar a}')$
is given by \eqEXdba.

\subsec{$\widehat{\frak{sl}_4}$, level-$1$}

The affine Lie algebra $\wh{\frak{sl}_4}$ at level $k=1$ has four 
integrable highest weight modules, corresponding to the singlet 
$\bf1$, the vector $L(\La_1)=\bf4$, the rank-2 
anti-symmetric tensor $L(\La_2)=\bf6$ and
the conjugate vector $L(\La_3)=\bar{\bf4}$.  
The UCPF form of the characters, corresponding to quasi-particles 
(intertwiners) transforming in the $\bf4$, $\bf6$ and $\bar{\bf4}$,
was obtained in [\BH].  To incorporate the null-states in the 
quasi-particle Fock space we need to deform both the inverse Cartan 
matrix of $\frak{sl}_4$
\eqn\eqEXfb{
\bB \eql \left( \matrix{ 
\thqu & \half & \onqu \cr
\half & \scr{1}   & \half \cr
\onqu & \half & \thqu \cr} \right)
}
as well as introduce one additional null-particle (corresponding to the
composite of two $\bf6$ particles).  The analogue of 
the $\frak{sl}_3$ expression \eqEXdf\ is given by \eqEXca\ with
\eqn\eqEXfa{ \eqalign{
& \bG \eql \cr
& \left( \matrix{
\thqu & \thqu & \thqu & \thqu & \vdots & \half & \half & \half & 
  \thha & \thha & \thha & \vdots & \onqu & \onqu & \onqu & \fiqu & 
\vdots & 2 \cr
\thqu & \thqu & \thqu & \thqu & \vdots & \half & \half & \half & 
  \half & \half & \thha & \vdots & \onqu & \onqu & \onqu & \onqu & 
\vdots & 2 \cr
\thqu & \thqu & \thqu & \thqu & \vdots & \half & \half & \half & 
  \half & \half & \half & \vdots & \onqu & \onqu & \onqu & \onqu & 
\vdots & 1 \cr
\thqu & \thqu & \thqu & \thqu & \vdots & \half & \half & \half & 
  \half & \half & \half & \vdots & \onqu & \onqu & \onqu & 
\onqu & \vdots & 1 \cr
\ldots&\ldots&\ldots&\ldots&\ldots&\ldots&\ldots&\ldots&\ldots&\ldots&
\ldots&\ldots&\ldots&\ldots&\ldots&\ldots&\ldots&\ldots\cr
\half & \half & \half & \half & \vdots & 1 & 1 & 1 & 1 & 1 & 2 &\vdots & 
\half & \half & \thha & \thha & \vdots & 2 \cr
\half & \half & \half & \half & \vdots & 1 & 1 & 1 & 1 & 2 & 1 &\vdots & 
\half & \half & \half & \thha & \vdots & 2 \cr
\half & \half & \half & \half & \vdots & 1 & 1 & 1 & 1 & 1 & 1 &\vdots & 
\half & \half & \half & \thha & \vdots & 2 \cr
\thha & \half & \half & \half & \vdots & 1 & 1 & 1 & 1 & 1 & 1 &\vdots & 
\half & \half & \half & \half & \vdots & 2 \cr
\thha & \half & \half & \half & \vdots & 1 & 2 & 1 & 1 & 1 & 1 &\vdots & 
\half & \half & \half & \half & \vdots & 2 \cr
\thha & \thha & \half & \half & \vdots & 2 & 1 & 1 & 1 & 1 & 1 &\vdots & 
\half & \half & \half & \half & \vdots & 2 \cr
\ldots&\ldots&\ldots&\ldots&\ldots&\ldots&\ldots&\ldots&\ldots&\ldots&
\ldots&\ldots&\ldots&\ldots&\ldots&\ldots&\ldots&\ldots\cr
\onqu & \onqu & \onqu & \onqu & \vdots & \half & \half & \half & \half & 
\half & \half & \vdots & \thqu & \thqu & \thqu & \thqu & \vdots & 1 \cr
\onqu & \onqu & \onqu & \onqu & \vdots & \half & \half & \half & \half & 
\half & \half & \vdots & \thqu & \thqu & \thqu & \thqu & \vdots & 1 \cr
\onqu & \onqu & \onqu & \onqu & \vdots & \thha & \half & \half & \half & 
\half & \half & \vdots & \thqu & \thqu & \thqu & \thqu & \vdots & 2 \cr
\fiqu & \onqu & \onqu & \onqu & \vdots & \thha & \thha & \thha & \half & 
\half & \half & \vdots & \thqu & \thqu & \thqu & \thqu & \vdots & 2 \cr
\ldots&\ldots&\ldots&\ldots&\ldots&\ldots&\ldots&\ldots&\ldots&\ldots&
\ldots&\ldots&\ldots&\ldots&\ldots&\ldots&\ldots&\ldots\cr
2 & 2 & 1 & 1 & \vdots & 2  & 2 & 2 & 2 & 2 & 2 & \vdots & 1 & 1 & 2 & 2 & 
\vdots & 4 \cr}
\right) \cr}
}
and $A_a=0$, $u_a=\infty$ and $\ta_a=1$ for all $a$.
The solution to \eqEXbf\ is given by
\eqn\eqEXfe{
\{\xi_a\} \eql \{ \txt{1\over7},\txt{2\over9},\txt{1\over4},\txt{1\over4}|
\txt{1\over10},\txt{1\over8},\txt{1\over7},\txt{1\over7},\txt{1\over8},
 \txt{1\over10}|\txt{1\over4},\txt{1\over4},\txt{2\over9},\txt{1\over7}|
\txt{1\over81} \} \,,
}
and leads to
\eqn\eqEXff{
c \eql \left( {6\over \pi^2}\right) \ \sum_a \ L(\xi_a) \eql 3\,,
}
as it should.  The solution of \eqEXbac\ with 
\eqn\eqEXfc{ \eqalign{
\{\la_a\} & \eql \{ \la_1,\la_2,\la_3,\la_4|\la_{12},\la_{13},\la_{14},
\la_{23},\la_{24},\la_{34}|\la_{123},\la_{124},\la_{134},\la_{234}|\mu\}\cr
\{ z_a \} & \eql \{\underbrace{x,\ldots,x}_4|\underbrace{x^2,\ldots,x^2}_6|
\underbrace{x^3,\ldots,x^3}_4|x^4\} \cr}
}
is given in Appendix A, and leads to 
\eqn\eqEXapb{ \eqalign{
\ze_1 & ~\equiv~ \left( \prod \la_i \right)^{3\over4}
                 \left( \prod \la_{ij} \right)^{1\over2}
          \left( \prod \la_{ijk} \right)^{1\over4} \mu\hphantom{^2}
     \eql 1 + 3x \,,\cr
\ze_2 & ~\equiv~ \left( \prod \la_i \right)^{1\over2}
                 \left( \prod \la_{ij} \right)\hphantom{^{1\over2}}
          \left( \prod \la_{ijk} \right)^{1\over2} \mu^2 \eql 1 + 2x + 3x^2 
   \,,\cr
\la_{\rm tot}^{1\over4} \eql 
\ze_3 & ~\equiv~ \left( \prod \la_i \right)^{1\over4}
                 \left( \prod \la_{ij} \right)^{1\over2}
    \left( \prod \la_{ijk} \right)^{3\over4} \mu\hphantom{^2}
   \eql 1 + x + x^2 + x^3 \,.
 \cr}
}

Note that the expression for $\latot$ is in complete agreement with
the results of [\Sca,\BSc], confirming that the exclusion statistics
of $(\wh{\frak{sl}_4})_{k=1}$ is indeed described by a 
statistical interaction matrix \eqEXfa,
while the $\ze_i(x=1)$ are in agreement with \eqEXqc.

\subsec{$\wh{\frak{sl}_n}$, level $k\geq1$}

Obtaining results for 
for $\frak{sl}_n$, $n\geq5$, at level-$1$, using the algorithm described
in [\BH] becomes extremely cumbersome.  No complete results are known,
but preliminary investigations suggest
\eqn\eqEXapd{ 
\ze_i(x) \eql \sum_{k=0}^i\ \bin{n-i-1-k}{k}\, x^k\,,\qquad 
\latot \eql \ze_n^{n-1}\,,
}
where $\ze_i$ is defined in \eqEXqb, such that indeed
\eqn\eqEXape{
\ze_i(1) \eql \sum_{k=0}^i\ \bin{n-i-1-k}{k} \eql \bin{n}{i} \eql
{\rm dim}\, L(\La_i)\,.
}
As explained in Section 4, once the 
results for the level $k=1$ UCPF characters
are known, one can immediately obtain the level $k>1$ characters by 
correcting for the level-$k$ pseudo-particles as in \eqEXlc.

\subsec{$\widehat{\frak{so}_5}$, level-$1$}

The affine Lie algebra $\wh{\frak{so}_5}$, at level $k=1$, has three
integrable highest weight representations, corresponding to the 
singlet ${\bf1}$, the vector $v= {\bf 5} = L(\La_1)$ and the spinor
$s={\bf4}=L(\La_2)$ of $\frak{so}_5$.
The UCPF form of the $(\wh{\frak{so}_5})_1$ characters is obtained 
by combining the results of [\Ya,\BSc,\BH].  In [\Ya] the 
character was given in terms of (restricted) $\frak{so}_5$ Kostka
polynomials and a recipe was given to compute the restricted Kostka
polynomial.  Explicit expressions for the restricted Kostka polynomial
(corresponding to the pseudo-particle part of the character) 
were given in [\BSc] while in [\BH] the UCPF form of the physical 
particles was found.  The final result requires one pseudo-particle,
physical particles transforming in the $\bf5$ and $\bf4$ of $\frak{so}_5$
and one null-particle (corresponding to the 
composite of two $\bf 5$ particles).  
The characters are given by \eqEXca\ with 
\eqn\eqEXea{
\bG \eql \left( \matrix{
1 & \vdots & 0 & 0 & 0 & 0 & 0 & \vdots & 
-\half & -\half & -\half & -\half  & \vdots & 0 \cr
\ldots&\ldots&\ldots&\ldots&\ldots&\ldots&\ldots&\ldots&\ldots&\ldots&
\ldots&\ldots&\ldots&\ldots\cr
0 & \vdots & 1 & 1 & 1 & 1 & 2 & \vdots & 
\half & \half & \thha & \thha & \vdots & 2 \cr
0 & \vdots & 1 & 1 & 1 & 2 & 1 & \vdots & 
\half & \half & \half & \thha & \vdots & 2 \cr
0 & \vdots & 1 & 1 & 1 & 1 & 1 & \vdots & 
\half & \half & \half & \thha & \vdots & 2 \cr
0 & \vdots & 1 & 2 & 1 & 1 & 1 & \vdots & 
\half & \half & \half & \half & \vdots & 2 \cr
0 & \vdots & 2 & 1 & 1 & 1 & 1 & \vdots & 
\half & \half & \half & \half & \vdots & 2 \cr
\ldots&\ldots&\ldots&\ldots&\ldots&\ldots&\ldots&\ldots&\ldots&\ldots&
\ldots&\ldots&\ldots&\ldots\cr
-\half & \vdots & \half & \half & \half & \half & \half & \vdots & 
\thqu & \thqu & 
  \thqu & \thqu & \vdots & 1 \cr
-\half & \vdots & \half & \half & \half & \half & \half & \vdots & 
\thqu & \thqu & 
  \thqu & \thqu  & \vdots & 1 \cr
-\half & \vdots & \thha & \half & \half & \half & \half & \vdots & 
\thqu & \thqu & 
  \thqu & \thqu & \vdots & 2 \cr
-\half & \vdots & \thha &\thha &\thha & \half & \half & \vdots & 
\thqu & 
  \thqu & \thqu & \thqu 
  & \vdots & 2 \cr
\ldots&\ldots&\ldots&\ldots&\ldots&\ldots&\ldots&\ldots&\ldots&\ldots&
\ldots&\ldots&\ldots&\ldots\cr
0 & \vdots & 2 &2&2&2&2&\vdots & 1&1&2&2&\vdots & 4 \cr}
\right) \,,
}
which is a deformation of the matrix 
\eqn\eqEXla{
\bB \eql \left( \matrix{ \scr{1} & \scr{1\over2} \cr 
                         \scr{1\over2} & \scr{3\over4} \cr} \right)\,,
}
entering \eqEXcb. Furthermore, $A_a=0$ and $\ta_a=1$ for all $a\in\cS$, and
\eqn\eqEXeba{ 
\bu  \eql \cases{ 
\{0|\underbrace{\infty,\ldots,\infty}_5|
\underbrace{\infty,\ldots,\infty}_4|\infty\} & for $1$ and $v\,,$ \cr
\{1|\underbrace{\infty,\ldots,\infty}_5|
\underbrace{\infty,\ldots,\infty}_4|\infty\} & for $s\,,$ \cr} 
}
while there are also some restrictions on the summation over 
the $m_a$'s (see [\BSc]).
Note that this case corresponds to order $k=2$ non-abelian exclusion
statistics in the sense of [\GS] as far as the coupling of the pseudo-particle
to the physical spinor-particles are concerned.
The physical vector-particles have a unique fusion rule and therefore do not
couple to the pseudo-particle.

The equations \eqEXbf\ have the solution
\eqn\eqEXeb{ \eqalign{
\{\xi_a\} & \eql \{ \scr{11\over16}\,|\, \scr{1\over12}, \scr{1\over8},
\scr{5\over33}, \scr{7\over40}, \scr{11\over60}\,|\, \scr{4\over11},
\scr{4\over11}, \scr{16\over49}, \scr{8\over33} \,|\, \scr{1\over49} \}\cr
\{\et_a\} & \eql \{ \half \,|\, 0,0,0,0,0\,|\,0,0,0,0\,|\,0 \}\cr}
}
leading to
\eqn\eqEXec{
c \eql \left( {6\over\pi^2}\right) \sum_a \big( L(\xi_a) - L(\et_a) \big)
\eql 3 - \txt{1\over2} \eql \txt{5\over2} \,,
}
as it should.  Moreover, we have verified that the total 1-particle
partition function $\la_{\rm tot} = \left(\prod_i \la_i \right)^2
\left(\prod_\al \la_\al \right) \mu^4$, resulting from the solution
of \eqEXbac\ with 
\eqn\eqEXecb{ \eqalign{
\{\la_a\} \eql & \{ \la|\la_1,\la_2,\la_0,\la_{\bar2},\la_{\bar1}|
\la_{++},\la_{+-},\la_{-+},\la_{--}|\mu\} \,,\cr
\{ z_a \} \eql & \{ 1 | \underbrace{x^2,\ldots,x^2}_5| \underbrace{x,\ldots,
x}_4 | x^4 \} \,, \cr}
}
satisfies,  up to at least $\cO(x^{11})$, the equation
\eqn\eqEXeca{
\latot^{3\over2} - (2+3x^2)\latot + (3x^2-1)(x^2-1)\latot^{1\over2}
- x^2(x^2-1)^2 \eql 0\,,
}
derived in [\BSb,\BSc] from the recursion approach (see Appendix B
for the explicit solution up to $\cO(x^{11})$).

In addition, from \eqEXeb, we obtain that the expressions for
\eqn\eqEXed{ \eqalign{
\la_{\rm tot}^{1\over2}  \eql \ze_1 \eql & \left(\prod \la_i \right)
\left(\prod \la_\al \right)^{1\over2} \mu^2\,, \cr
\ze_2  \eql & \la^{1\over2} \left(\prod \la_i \right)^{1\over2}
\left(\prod \la_\al \right)^{1\over4} \mu \,,\cr}
}
at $x=1$ are given by, respectively, $\ze_1=5$ and $\ze_2=4$.  Again 
this is in complete agreement with \eqEXqc.
The results in this section might prove to be useful with regards to
certain quasi-particle excitations (`non-abelian electrons') in
$SO(5)$ superspin regimes for strongly correlated electrons on a 
two-leg ladder [\BSb].

The UCPF and corresponding exclusion statistics for higher level
$\wh{\frak{so}_5}$ modules can be worked out using the results of 
[\Ya].

\subsec{$\wh{\frak{sl}_n}$, level-$1$, revisited}

For $(\wh{\frak{sl}_n})_{k=1}$ it is also possible to give a description 
purely in terms of quasi-particles (`spinons') $\ph^a$ transforming in the
$n$-dimensional vector representation $\bf n$. 
In this case the null-field will be of the form $:\!\ph^1(z) \ldots
\ph^n(z)\!:$.  The corresponding character formula was found in [\BSa]
\eqn\eqEXga{
{\rm ch}_i(y;q) \eql \sum_{m_a \geq 0 \atop \sum m_a \equiv i\,
{\rm mod}\, n} y^{\sum m_a} 
\sum_{m\geq0} (-1)^m {q^{ {1\over2}m(m-1)}\over \qn{m}}
{q^{ {1\over2} ( \sum m_a^2 - {1\over n} (\sum m_a)^2 )} \over
\prod_a \qn{m_a-m} }\,.
}
It can be brought in the UCPF form by shifting the summation variables
$m_a \to m_a +m$ (for $y\neq1$).  Then,
\eqn\eqEXgb{
\bG \eql \left( \matrix{ 
  & & & \vdots & \cr
 & \de_{ab} - \txt{1\over n} & & \vdots & 0 \cr
  & & & \vdots & \cr
\ldots & \ldots & \ldots & \ldots & \ldots \cr
 & 0 & & \vdots & 1 \cr} \right) \,.
}
The corresponding equations \eqEXbac\ 
with $\{\la_a\} = \{ \la_1,\ldots,\la_n | 
\mu\}$, and $\{z_a\} = \{ x,\ldots,x|-x^n\}$ have the solution
\eqn\eqEXgc{
\la_1 \eql \ldots \eql \la_n \eql {1\over 1-x} \,,\qquad 
\mu \eql 1-x^n\,,
}
so that indeed
\eqn\eqEXgd{
\la_{\rm tot} \eql (\la_1 \ldots\la_n) \mu^n \eql (1+x+\ldots+x^{n-1})^n\,.
}

It is also possible to write \eqEXga\ in terms of a non-alternating 
sum by repeated application of \eqEXcg\ and \eqEXch.  Besides the 
$n$ spinons this requires $n-2$ additional null-particles for $\frak{sl}_n$. 
Here give the result 
for $\wh{\frak{sl}_3}$ (see [\BH] for the origin of this 
formula and the generalization to $\wh{\frak{sl}_n}$)
\eqn\eqEXge{
{\rm ch}_i(y;q) \eql \sum_{m_a,m\geq0} y^{\sum m_a + 2m} 
{q^{{1\over2} \bm\cdot \bG \cdot \bm} \over \qn{m_1}\qn{m_2}\qn{m_3}\qn{m}}
\,,}
with 
\eqn\eqEXgf{ 
\bG \eql \left( \matrix{ 
\twth    & \twth   & -\onth   & \vdots  & \onth  \cr
\twth    & \twth   & -\onth   & \vdots  & \onth  \cr
-\onth   & -\onth  & \twth    & \vdots  & \onth  \cr
\ldots   & \ldots  & \ldots   & \ldots  & \ldots \cr
\onth    & \onth   & \onth    & \vdots  & \twth  \cr} \right)\,,
}
leading to the solution
\eqn\eqEXgg{
\la_1 \eql \la_2 \eql 1+x\,, \qquad \la_3 \eql 1 + x + x^2 \,,
\qquad \mu \eql {1+x+x^2\over 1+x} \,,
}
and again confirming
\eqn\eqEXgh{
\la_{\rm tot} \eql (\la_1\la_2\la_3)\mu^2 \eql (1 + x + x^2)^3\,.
}
In contrast to the UCPF formulas in section 4.2 and 4.3, it does not 
appear that the formula \eqEXga\ has a straightforward generalization
to levels $k>1$.

\newsec{Conclusions}

In this work we have to tried to reconcile Haldane's notion of exclusion
statistics [\Ha] with the Stony Brook group's proposal of a Universal Chiral
Partition Function form for all (chiral) characters of two-dimensional
conformal field theories [\BM].  We have seen that besides the
pseudo-particles of [\GS], in general, this requires yet another kind
of particles, so-called null-particles.  In support of the conjectured
relation between Haldane statistics and the UCPF, we have shown that
an ideal gas of physical, pseudo- and null-particles, with linear
dispersion relations, in the thermodynamic limit exhibits the same
effective central charge as the UCPF. It
would of course be most desirable to extend this comparison to the
different sectors of the UCPF and gain an understanding of the
restrictions that enter the sum.  

The UCPF was put forward to structuralize the form of the characters
of CFT.  By indicating how the characters of affine Lie algebras may
be written in the UCPF form by introducing null-particles we have
obtained further support for the alleged `universality' of the UCPF.

To demonstrate this method we
have discussed various examples of UCPFs for WZW-models and the
associated exclusion statistics and found agreement with previous
results, computed by the recursion method [\Sca,\BSc], in all cases.

\bigskip
\leftline{\bf Acknowledgements}\medskip

We would like to thank Sathya Guruswamy and Kareljan Schoutens for
discussions, useful remarks, and for making their manuscript [\GS] available
to us prior to publication.  P.B.\ is supported by a \qeii\ research
fellowship from the Australian Research Council and D.R.\ was
supported by a University of Adelaide Faculty of Science summer
scholarship.

\vfil\eject

\appendix{A}{Solution for $\frak{sl}_4$}

The explicit solution of \eqEXbac\  with \eqEXfc\ is given by
\eqn\EXapa{ \eqalign{
\la_1 &    \eql {1+3x+3x^2\over 1+2x+3x^2 } 
           \eql 1 + {x\over 1+2x+3x^2 } \,,\cr
\la_2 &    \eql {(1+2x)^2\over 1+3x+3x^2 }
           \eql 1+{x(1+x)\over 1+3x+3x^2 }\,,\cr
\la_3 &    \eql {1+3x\over1+2x} 
           \eql 1 + {x\over 1+2x} \,,\cr
\la_4 &    \eql {1+3x\over1+2x} 
           \eql 1 + {x\over 1+2x} \,,\cr
\la_{12} & \eql {(1+2x+2x^2)(1+3x+2x^2+2x^3)\over(1+2x)^2(1+x+x^2+x^3)} 
           \eql {x^2(1+3x)\over (1+2x)^2(1+x+x^2+x^3)}\,,\cr
\la_{13}&  \eql {(1+x)^2(1+2x+3x^2)\over (1+x+x^2)(1+3x+3x^2)} 
           \eql 1+{x^2(1+2x) \over (1+x+x^2)(1+3x+3x^2)} \,,\cr
\la_{14}&  \eql {1+2x+3x^2+x^3\over (1+x)(1+x+x^2)}
           \eql 1+{x^2\over (1+x)(1+x+x^2)}\,,\cr
\la_{23}&  \eql {1+3x+3x^2\over (1+x)(1+2x)} 
           \eql 1+ {x^2\over (1+x)(1+2x)}\,,\cr
\la_{24}&  \eql {(1+x)^2(1+2x+3x^2)\over (1+2x)(1+2x+3x^2+x^3)}
           \eql 1+{x^2(1+x+x^2)\over (1+2x)(1+2x+3x^2+x^3)}\,,\cr
\la_{34}&  \eql {(1+2x+2x^2)(1+3x+2x^2+2x^3)\over (1+3x)(1+x+x^2)^2}
           \eql 1+{x^2(1+x+x^2+x^3) \over (1+3x)(1+x+x^2)^2}\,,\cr
\la_{123}& \eql  {1+x+x^2+x^3\over 1+x+x^2}
           \eql 1+{x^3\over 1+x+x^2}\,,\cr
\la_{124} &\eql {1+x+x^2+x^3\over 1+x+x^2}
           \eql 1+{x^3\over 1+x+x^2}\,,\cr
\la_{134}& \eql {(1+x+x^2)^2\over 1+2x+3x^2+x^3}
           \eql 1+{x^3(1+x)\over  1+2x+3x^2+x^3}\,,\cr
\la_{234}& \eql {1+2x+3x^2+x^3\over 1+2x+3x^2}
           \eql 1+{x^3\over 1+2x+3x^2 } \,,\cr
\mu      & \eql {(1+2x)^2(1+x+x^2)^2\over (1+x)(1+2x+2x^2)(1+3x+2x^2+2x^3)}\cr
         &  \eql 1 + {x^4\over (1+x)(1+2x+2x^2)(1+3x+2x^2+2x^3)}
\,,\cr}
}

\vfil\eject

\appendix{B}{Approximate solution for $\frak{so}_5$}

Up to $\cO(x^{11})$ the solution of \eqEXbac\  with \eqEXecb\ is given by
\eqn\eqEXapba{\eqalign{
\la \eql &  2+4y-8y^2+18y^3-48y^4+\txt{303\over2}y^5-544y^6+
        \txt{8505\over4}y^7-8768y^8 +\txt{1198427\over32}y^9\cr & -
        163968y^{10}+\cO(y^{11}) \,, \cr
\la_1 \eql &  1+2y^2-16y^3+108y^4-696y^5+4408y^6-27702y^7+
        173424y^8-1083451y^9\cr & +6760800y^{10}+ \cO(y^{11}) \,, \cr
\la_2 \eql & 1+2y^2-12y^3+64y^4-334y^5+1736y^6-\txt{18053\over2}y^7+
        47008y^8-\txt{980991\over4}y^9\cr & 
        +1281696y^{10}+\cO(y^{11}) \,, \cr
\la_0 \eql &  1+2y^2-12y^3+68y^4-374y^5+2024y^6-\txt{21709\over2}y^7+
        57904y^8-\txt{1231627\over4}y^9\cr&+1634080y^{10}+\cO(y^{11}) \,, \cr
\la_{\bar2} \eql &  1+2y^2-8y^3+28y^4-92y^5+272y^6-619y^7+
        160y^8+\txt{21685\over2}y^9-100320y^{10}\cr&+\cO(y^{11}) \,, \cr
\la_{\bar1} \eql &  1+2y^2-8y^3+28y^4-84y^5+152y^6+569y^7-
        9616y^8+\txt{166483\over2}y^9-601248y^{10}\cr&+\cO(y^{11}) \,, \cr
\la_{++} \eql &  1+2y-6y^2+25y^3-116y^4+\txt{2255\over4}y^5-2808y^6+
        \txt{113577\over8}y^7-72496y^8\cr&+\txt{23858843\over64}y^9-
        1926944y^{10}+\cO(y^{11}) \,, \cr
\la_{+-} \eql &  1+2y-6y^2+25y^3-116y^4+\txt{2255\over4}y^5-2808y^6+
        \txt{113577\over8}y^7-72496y^8\cr&+\txt{23858843\over64}y^9-
        1926944y^{10}+\cO(y^{11}) \,, \cr
\la_{-+} \eql &  1+2y-6y^2+21y^3-80y^4+\txt{1255\over4}y^5-1240y^6+
        \txt{39117\over8}y^7-19104y^8+\txt{4700235\over64}y^9\cr & -
        275296y^{10}+\cO(y^{11}) \,, \cr
\la_{--} \eql &  1+2y-6y^2+13y^3-24y^4+\txt{151\over4}y^5-40y^6-
        \txt{363\over8}y^7+576y^8-\txt{203605\over64}y^9 \cr & 
        +15264y^{10}+\cO(y^{11}) \,, \cr
\mu \eql &  1+4y^4-48y^5+400y^6-2872y^7+19072y^8-
        120906y^9+743936y^{10}+\cO(y^{11}) \,, 
\cr}
}
where $y= x/\sqrt{2}$.  This leads to 
\eqn\eqEXapbb{ \eqalign{
\latot^{1\over2} \eql & \big( \prod_i \la_i \big) \big( 
\prod_\al \la_\al \big)^{1\over2} \mu \eql 
1+4y+2y^2+2y^3-8y^4+\txt{63\over2}y^5-128y^6
\cr & + \txt{2145\over4}y^7-2304y^8
+\txt{323323\over32}y^9-45056y^{10} + \cO(y^{11}) \,. \cr}
}


\footatend\vfill\eject\immediate\closeout\rfile
\baselineskip=14pt{{\bf  References}}\bigskip{\frenchspacing%
\parindent=20pt\escapechar=` \input refs.tmp\vfill\eject}\nonfrenchspacing
\vfill\eject\end